\documentclass[conference]{IEEEtran}
\IEEEoverridecommandlockouts

\usepackage{xcolor}
\usepackage{listings}
\usepackage[most]{tcolorbox}

\definecolor{codebg}{RGB}{248,248,248}
\definecolor{codeframe}{RGB}{220,220,220}
\definecolor{commentgray}{RGB}{100,100,100}
\definecolor{keywordblue}{RGB}{0,92,197}
\definecolor{stringred}{RGB}{163,21,21}

\lstdefinestyle{pythonstyle}{
  language=Python,
  basicstyle=\ttfamily\scriptsize,
  keywordstyle=\color{keywordblue}\bfseries,
  commentstyle=\color{commentgray}\itshape,
  stringstyle=\color{stringred},
  numbers=left,
  numberstyle=\tiny\color{gray},
  stepnumber=1,
  numbersep=7pt,
  showstringspaces=false,
  breaklines=true,
  breakatwhitespace=true,
  tabsize=4,
  keepspaces=true,
  columns=fullflexible,
  frame=none,
  xleftmargin=0pt,
  aboveskip=0pt,
  belowskip=0pt
}

\newtcblisting{pythonbox}{
  listing only,
  listing options={style=pythonstyle},
  colback=codebg,
  colframe=codeframe,
  boxrule=0.4pt,
  arc=2pt,
  left=6pt,
  right=6pt,
  top=5pt,
  bottom=5pt,
  enhanced,
  sharp corners=south,
}

\usepackage{booktabs}
\usepackage{enumitem}
\usepackage{array}
\usepackage{wrapfig}
\usepackage{amsmath,amsfonts}
\usepackage[noend]{algpseudocode}
\usepackage{graphicx}
\usepackage{textcomp}
\usepackage{float}
\usepackage{listings}
\usepackage{xspace}
\usepackage{multirow}
\usepackage{amsthm}

\usepackage{balance}
\usepackage{algorithm}
\usepackage{algpseudocode}
\usepackage{colortbl}

\usepackage{multicol}
\newcommand*\colourcheck[1]{%
	\expandafter\newcommand\csname #1check\endcsname{\textcolor{#1}{\ding{52}}}%
}
\colourcheck{blue}
\colourcheck{green}
\colourcheck{red}
\newtcolorbox{boxB}[2][]{%
  enhanced,colback=white,colframe=black,coltitle=black,
  sharp corners,
  toprule=1.0pt,
  rightrule=0.3pt,
  leftrule=0pt,
  bottomrule=0pt,
  fonttitle=\itshape\scshape\large,
  left=0pt,right=5pt,top=5pt,bottom=3pt,
  attach boxed title to top right={yshift=-0.3\baselineskip-0.4pt,xshift=-5mm},
  boxed title style={tile,size=minimal,left=0.2mm,right=0.5mm,
    colback=white,before upper=\strut},
  title=#2,#1
}

\newcommand{\tool}{\textsc{SpecCoder}\xspace}
\newcommand{\posdelta}[1]{\textcolor{deepgreen}{+#1\%}}
\definecolor{deepred}{RGB}{180, 35, 35}
\newcommand{\negdelta}[1]{\textcolor{deepred}{#1\%}}

\def\BibTeX{{\rm B\kern-.05em{\sc i\kern-.025em b}\kern-.08em
    T\kern-.1667em\lower.7ex\hbox{E}\kern-.125emX}}

\newboolean{showcomments}
\setboolean{showcomments}{true}
\ifthenelse{\boolean{showcomments}}
 { \newcommand{\mynote}[2]{
      \fbox{\bfseries\sffamily\scriptsize#1}
        {\small$\blacktriangleright$\textsf{\emph{#2}}$\blacktriangleleft$}}}
        { \newcommand{\mynote}[2]{}}

\usepackage{tikz}

\definecolor{neutralgray}{RGB}{110,110,110} 
\definecolor{deepgreen}{RGB}{0,120,0} 
\newcommand{\basedelta}{\textcolor{neutralgray}{0.0\%}} 

\newcolumntype{L}[1]{>{\raggedright\arraybackslash}p{#1}}

\newcommand{\code}[1]{{\footnotesize\texttt{#1}}}
\usepackage{amsthm}
\definecolor{dkgreen}{rgb}{0,0.6,0}
\definecolor{gray}{rgb}{0.5,0.5,0.5}
\definecolor{lightgray}{rgb}{211, 211, 211}
\definecolor{mauve}{rgb}{0.58,0,0.82}
\definecolor{lightblue}{RGB}{245, 245, 220}
\definecolor{grey}{RGB}{169,169,169}
\definecolor{darkgreen}{RGB}{54, 156, 90}
\definecolor{lightblue}{RGB}{235, 247, 255}
\definecolor{niceblue}{RGB}{52, 164, 235}
\definecolor{nicered}{RGB}{219, 42, 48}
\definecolor{niceorange}{RGB}{224, 114, 54}
\definecolor{lightred}{RGB}{255, 227, 228}
\definecolor{lightorange}{RGB}{252, 235, 227}
\definecolor{orange1}{RGB}{214, 178, 131}
\definecolor{brown}{RGB}{163, 117, 57}
\definecolor{lightbrown}{RGB}{224, 163, 83}
\definecolor{commentcolor}{RGB}{101, 163, 178}
\definecolor{lightpurple}{RGB}{203, 195, 227}
\newcolumntype{L}[1]{>{\raggedright\arraybackslash}p{#1}}
\usepackage{amsthm}
\definecolor{dkgreen}{rgb}{0,0.6,0}
\definecolor{deepgreen}{rgb}{0,0.45,0}
\definecolor{gray}{rgb}{0.5,0.5,0.5}
\definecolor{lightgray}{rgb}{211, 211, 211}
\definecolor{mauve}{rgb}{0.58,0,0.82}

\definecolor{custom-blue}{rgb}{0,0,0}
\definecolor{custom-red}{rgb}{1,0,0}

\definecolor{c1}{HTML}{f4cccc}
\definecolor{c2}{HTML}{f5cdcd}
\definecolor{c3}{HTML}{fffcfc}
\definecolor{c4}{HTML}{ffffff}
\definecolor{c5}{HTML}{ffffff}
\definecolor{c6}{HTML}{fffdfd}
\definecolor{c7}{HTML}{f5cfcf}
\definecolor{c8}{HTML}{fffbfb}
\definecolor{c9}{HTML}{ffffff}
\definecolor{c10}{HTML}{fffdfd}
\definecolor{c11}{HTML}{fefafa}
\definecolor{c12}{HTML}{fef7f7}
\definecolor{c13}{HTML}{ffffff}
\definecolor{c14}{HTML}{fffefe}
\definecolor{c15}{HTML}{ffffff}
\definecolor{c16}{HTML}{fefafa}
\definecolor{c17}{HTML}{fdf3f3}
\definecolor{c18}{HTML}{fffefe}
\definecolor{c19}{HTML}{fdf5f5}
\definecolor{c20}{HTML}{ffffff}
\definecolor{natboundary}{RGB}{64,105,215}      
\definecolor{formalboundary}{RGB}{205,92,92}    
\definecolor{checkpointgreen}{RGB}{60,150,115}  

\makeatletter
\newcommand{\linebreakand}{%
  \end{@IEEEauthorhalign}
  \hfill\mbox{}\par
  \mbox{}\hfill\begin{@IEEEauthorhalign}
}
\makeatother

\begin{document}

\title{Teaching Code LLMs to Reason with Intermediate Formal Specifications}


\author{
\IEEEauthorblockN{Minh Le-Anh}
\IEEEauthorblockA{\textit{FPT Software AI Center} \\
\textit{Hanoi Univ. of Science and Tech.} \\
Hanoi, Vietnam \\
minhla4@fpt.com}
\and
\IEEEauthorblockN{Cuong Chi Le}
\IEEEauthorblockA{\textit{University of Texas at Dallas} \\
Texas, USA \\
cuong.le@utdallas.edu}
\and
\IEEEauthorblockN{Tien N. Nguyen}
\IEEEauthorblockA{\textit{University of Texas at Dallas} \\
Texas, USA \\
tien.n.nguyen@utdallas.edu}
}

\maketitle

\begin{abstract}

Unlike natural-language specifications, executable formal specifications provide
machine-checkable constraints for verifying, debugging, and repairing code.
However, writing such specifications is labor-intensive, and existing LLM-based
methods mainly infer whole-program pre/postconditions, missing the intermediate
semantic commitments that programmers rely on when reasoning about an algorithm.
Our study further shows that prompting current CodeLLMs often produces executable
assertions that are syntactically invalid, trivial, or too weak to reject
behavior-changing faults. In this paper, we study executable checkpoint specification generation, where assertions are inserted at meaningful internal program points to describe
expected intermediate states. We introduce {\tool}, a verification-guided
CodeLLM training framework that learns from validated reference programs,
behavior-changing mutants, and multi-turn specification-refinement traces.
{\tool} selects specifications that hold on correct executions while rejecting
faulty executions, turning specifications from passive annotations into
executable evidence. To evaluate this setting, we introduce HumanExec, a benchmark built from recent Codeforces competitive-programming problems with test suites, reference
solutions, and human buggy submissions, supporting three tasks: specification
generation, program correctness checking, and program repair. Experiments on
HumanExec show that {\tool} substantially improves checkpoint-specification
quality over base CodeLLMs. Across Qwen2.5-Coder models, {\tool} improves
inline-specification correctness by up to {\bf 55.8\%}, completeness by up to
{\bf 358.1\%}, and executable assertion validity by up to {\bf 26.6\%}. These
gains further translate to downstream correctness reasoning and repair, showing
that executable checkpoints provide fine-grained evidence for reliable
verification.

\end{abstract}


\section{Introduction}
\label{sec:intro}

Formal specifications are symbolic constraints that describe expected
program behavior and can be checked against executions~\cite{BeschastnikhBSSE2011,daikon99}. When a specification is written as an assertion, it becomes directly executable:
the assertion can be evaluated on concrete program traces and used as
feedback for verification, testing, debugging, and repair. 
Instead of only asking whether a program produces the correct final output, executable
specifications can expose whether the program satisfies the properties
needed to reach that output correctly.

Recent advances in Large Language Models (LLMs) have transformed software
development.
However, the hallucination and semantic unreliability of LLM-generated code have shifted the focus
of software engineering research from merely generating code to a more
fundamental question: how can we verify that generated code is correct
and aligned with user intent, and how can we help repair it when it is
not? Program specifications provide an important vehicle for this goal.
They make the intended behavior explicit and can serve as oracles for
checking, explaining, and correcting generated implementations.
Unfortunately, manually writing high-quality formal specifications for
LLM-generated code is labor-intensive and requires substantial
programming and verification expertise. This creates a clear need for
automated techniques that can generate executable formal specifications.

Recent research has shown that LLMs can generate program specifications
for a given implementation, including postconditions and natural-language
correctness conditions~\cite{nl2postcond}. For example, NL2Postcond~\cite{nl2postcond} and SpecMind~\cite{le-etal-2026-specmind} infer postconditions from source code. HoarePrompt~\cite{hoareprompt} uses natural-language state conditions for correctness reasoning, and SpecRover~\cite{ruan2025specrover} extracts code intent to support repair. They demonstrate the value of specification-like reasoning, but their intermediate signals are not directly executable assertions.


First, most existing approaches derive specifications at a coarse
granularity. Approaches such as NL2Postcond and SpecRover primarily focus
on preconditions or postconditions for an entire method or program.
End-to-end specifications are useful because they summarize boundary
behavior: what must hold before execution and what should hold after the
program terminates. However, they do not expose whether the main
algorithm is correct at each internal step that could ensure the final correctness. In practice, developers
rarely reason about non-trivial programs only through final outputs.
Instead, they decompose a program into meaningful computation steps and
check local semantic commitments: after preprocessing, an array should
have a certain structure; after a loop iteration, an invariant should
hold; etc. Whole-program postconditions often miss these
internal obligations, making them insufficient for fine-grained
verification, diagnosis, and repair.

Second, some LLM-based approaches can produce intermediate reasoning
conditions, but these conditions are often expressed in natural language.
For example, HoarePrompt-style reasoning can describe reachable program
states at intermediate points, helping an LLM reason about correctness.
However, natural-language specifications are difficult to execute
directly. They may be useful for explanation or human inspection, but
they provide limited support for fully automated workflows that require
machine-checkable feedback, such as trace-based consistency checking,
test-time assertion validation, fault localization, and automated program
repair. 



These limitations motivate a richer view of specification for
LLM-generated code. Unlike prior work that treats specifications mainly as final summaries or textual reasoning aids, our goal is to {\em teach models to produce the executable intermediate specifications that are validated by execution and selected for their ability to reject faulty behaviors}.

\[
\textcolor{formalboundary}{\{\mathsf{Pre}\}}\; S_0\;
\textcolor{checkpointgreen}{\{\mathsf{Step}_1\}}\; S_1\;\cdots\;
\textcolor{checkpointgreen}{\{\mathsf{Step}_n\}}\; S_n\;
\textcolor{formalboundary}{\{\mathsf{Post}\}} .
\]
In this paper, we focus on assertion-style executable formal specifications: Boolean predicates inserted into source code and evaluated over concrete runtime states.
Such specifications still describe the
boundary behavior of the code, but they are executable constraints. The
\textcolor{checkpointgreen}{intermediate formal checkpoint
specifications} describe the semantic states reached by important
algorithmic steps. Given a program, these checkpoint specifications are executable
assertions placed inside the source code, so they can verify not only the
final answer but also the path by which the program reaches it. 


Toward this goal, our preliminary study shows that LLMs do not produce high-quality specifications in the intermediate checkpoints for a given program.
A final postcondition mainly requires
understanding the desired output behavior of the whole program, whereas a
checkpoint specification requires understanding the internal algorithm,
identifying meaningful program points, and expressing properties that
should hold at those points. Thus, checkpoint specification generation
demands a deeper form of step-by-step program reasoning. The gap exposed
by our study suggests that current code LLMs may produce plausible final
conditions while still lacking robust understanding of intermediate
program execution. 
Moreover, they often produce trivial assertions,
or weak conditions that are valid but provide little
help for verification.



We introduce {\tool}, a specification-aware training framework that teaches code LLMs to generate executable checkpoint specifications for a given program. {\tool} constructs training data from correct programs, synthesized mutants, and specification-refinement traces, then trains models to produce interleaved code-and-specification outputs. The resulting specifications are executable assertions that can be evaluated on program traces for correctness. The trained/fine-tuned model supports the annotation setting (specification generation): given a program, it inserts executable checkpoint specifications. This paper makes the following contributions:

\begin{itemize}[leftmargin=*, nosep]

    \item A verification-guided training framework that uses reference
    programs, validated mutants, and specification-refinement traces to
    produce specification-aware supervision for LLMs.

    \item HumanExec, an execution-based benchmark for assessing
    specification generation on tasks with human-written buggy code, measuring both correctness and completeness.


    \item We empirically study the usefulness of specification-aware
    training in three downstream applications: {\bf \em specification generation}, {\bf \em consistency checking} between program intent
    and implementation, and {\bf \em program repair}.
\end{itemize}

\section{Motivation}
\label{sec:motivation}

\subsection{Limitations of Natural-Language Specifications for Automated Verification}

Natural-language specifications are widely used in recent LLM-based verification and program-repair methods because they are easy to generate, read, and communicate to developers~\cite{hoareprompt}. For example, an LLM can describe an intermediate state with a sentence such as ``the list should be sorted before merging''. Such descriptions are useful for human inspection and can guide an LLM's reasoning about code.

However, the key limitation is that natural-language specifications are not executable. Unlike formal assertion-style specifications, they do not define precise Boolean conditions over concrete program states. Therefore, they cannot be directly checked against execution traces. 



Fig.~\ref{fig:motiv-case} illustrates this with a simplified interval-merging program. The natural-language checkpoint states that the intervals should be sorted before merging. This description is readable, but it cannot be evaluated automatically. If the implementation accidentally sorts intervals by their end point instead of their start point, an automated system cannot directly detect the violation. In contrast, the executable assertion at line 13 directly checks the required intermediate state on the concrete runtime value of \code{intervals}.

\definecolor{githubgreen}{RGB}{26,127,55}
\definecolor{githublight}{RGB}{230,255,237}
\definecolor{githubred}{RGB}{220,53,69}
\definecolor{githubredlight}{RGB}{255,235,238}

\lstset{
  basicstyle=\ttfamily\scriptsize,
  numbers=left,
  numberstyle=\tiny,
  breaklines=true,
  xleftmargin= 10pt,
}

\begin{figure}[t]
\begin{lstlisting}[language=Python]
def merge_intervals(intervals):
# BUG: intervals should be sorted by start point,
# but this implementation sorts by end point.
# Example input: [[10,20], [1,100]]
# Sorting by end keeps [10,20] before [1,100], so the start points are not sorted.
intervals.sort(key=lambda p: p[1])

# Natural-language spec at the checkpoint:
# "The intervals should be sorted before merging."
# This is readable, but cannot be checked automatically.

# Executable checkpoint specification:
assert all(intervals[i][0] <= intervals[i+1][0]
           for i in range(len(intervals) - 1))

merged = []
for start, end in intervals:
    if not merged or merged[-1][1] < start:
        merged.append([start, end])
    else:
        merged[-1][1] = max(merged[-1][1], end)

return merged
\end{lstlisting}
\vspace{-10pt}
\caption{Natural-language versus Executable Checkpoint Specification}
\label{fig:motiv-case}
\end{figure}






Natural-language specifications can also be vague or incomplete. For example, an LLM may state that ``the current interval has been handled correctly'' or ``the merged list is updated properly.'' These statements sound plausible, but they do not specify what must hold over program variables, such as whether the intervals are ordered, or whether the last interval contains the maximum end point seen so far.


This limitation weakens downstream automation. In program verification, a natural-language specification cannot be evaluated during execution. In consistency checking, it cannot provide an objective signal that an implementation violates the user intent. In program repair, it provides only weak feedback: the system may know that a final test failed, but not which intermediate semantic commitment was broken. In contrast, an executable checkpoint can expose the internal state where the computation first deviates from the intended behavior. 

This motivates executable checkpoint specifications: assertions placed at meaningful program points that can be validated on executions and used as reliable verification signals.

\subsection{Limitations of Code LLMs for Generating Reliable Formal Specifications}

Executable checkpoint specifications address the non-verifiability of natural-language descriptions, but generating them is substantially harder than generating free-form text. A model must identify meaningful program points, infer the intended semantics of the intermediate state, determine which variables are in scope, and express the property as a syntactically valid Boolean predicate. A generated assertion that refers to an unavailable variable, fails on the correct program, or states a tautology such as \code{assert True} cannot serve as a reliable verification signal.

We thus examine whether base Code LLMs can generate reliable inline formal specifications through prompting alone. For each model, we prompt it to insert checkpoint assertions into correct programs and evaluate the generated assertions from three execution-based perspectives. \emph{Syntax} measures whether the assertions are executable. \emph{Correctness} measures whether they hold on correct executions. In our mutant-based evaluation, \emph{Completeness} measures, operationally, whether they reject behavior-changing mutants that reach the same checkpoints and therefore provide useful verification signals.

For example, for the program in Fig.~\ref{fig:motiv-case}, a prompted model may generate a syntactically valid but weak assertion such as \code{assert len(intervals) >= 0}, which holds on both correct and faulty executions. It may also generate an overly strong assertion such as checking that intervals are already non-overlapping before the merge loop, which can fail on correct inputs. These illustrate why executable syntax alone is not sufficient: a useful checkpoint must be both valid for the intended computation and discriminative against faulty behavior.

Table~\ref{tab:motivation-base-spec-quality} shows that prompting alone is insufficient for~reliable formal checkpoint specification generation. Although the models often produce syntactically executable assertions, their semantic quality remains limited. For example, Qwen2.5-Coder-32B achieves 0.89 syntax validity, but only 0.63 correctness and 0.26 completeness. This indicates that many generated assertions either fail on correct executions or hold on correct executions, yet are too weak to reject buggy variants. 

These results reveal a key challenge: executable assertions are not useful merely because they are well formed. They must also capture semantic properties that are valid for the intended computation and discriminative against faulty behavior. LLMs often miss this balance, producing assertions that are invalid, overly specific, tautological, or unrelated to the algorithmic state. This motivates a verification-guided training approach, where candidate assertions are refined using execution feedback and chosen for both correctness and discriminative power.



\begin{table}[t]
\centering
\small
\caption{Formal specification quality of base models before fine-tuning.}
\label{tab:motivation-base-spec-quality}
\resizebox{\columnwidth}{!}{%
\begin{tabular}{@{}lccc@{}}
\toprule
\textbf{Model}  & \textbf{Syntax} $\uparrow$ & \textbf{Correctness} $\uparrow$ & \textbf{Completeness} $\uparrow$ \\
\midrule
Qwen2.5-Coder-7B  & 0.8272 & 0.2772 & 0.1707 \\
Qwen2.5-Coder-14B & 0.7900 & 0.5226 & 0.1989 \\
Qwen2.5-Coder-32B & 0.8989 & 0.6313 & 0.2604 \\
\bottomrule
\end{tabular}%
}
\end{table}

\subsection{Key Idea}
\label{sec:speccoder-key-idea}

These observations suggest that checkpoint specification generation should not be treated as a purely textual generation problem. A useful checkpoint must be validated on correct executions and challenged against faulty executions. This motivates our key idea to turn specification generation into
verification-guided learning. Instead of treating an assertion as useful because
it appears plausible to an LLM, we evaluate candidate assertions through
execution. A checkpoint specification is useful only when it satisfies two
properties: it should hold on correct executions, and it should reject faulty
executions that reach the same checkpoint.

Given a programming problem \(x\) and an implementation \(y\), checkpoint
specification generation produces an annotated program \(\tilde{y}\), where
assertions are inserted at internal program points. Each assertion \(s\) is a
Boolean predicate over variables in scope at its insertion point. We evaluate
these assertions on two execution populations: \emph{correct executions}, where
the assertions should hold, and \emph{mutant executions}, where strong
assertions should expose behavior-changing faults. This converts specification
generation from a purely generative task into a feedback-driven process in
which the quality of a specification can be checked by running the program.


\section{{\tool}: A Verification-Based Model}
\label{sec:speccoder}

\subsection{Data Construction}
\label{sec:speccoder-data-construction}

\begin{figure*}[t]
\centering
\includegraphics[width=0.9\linewidth]{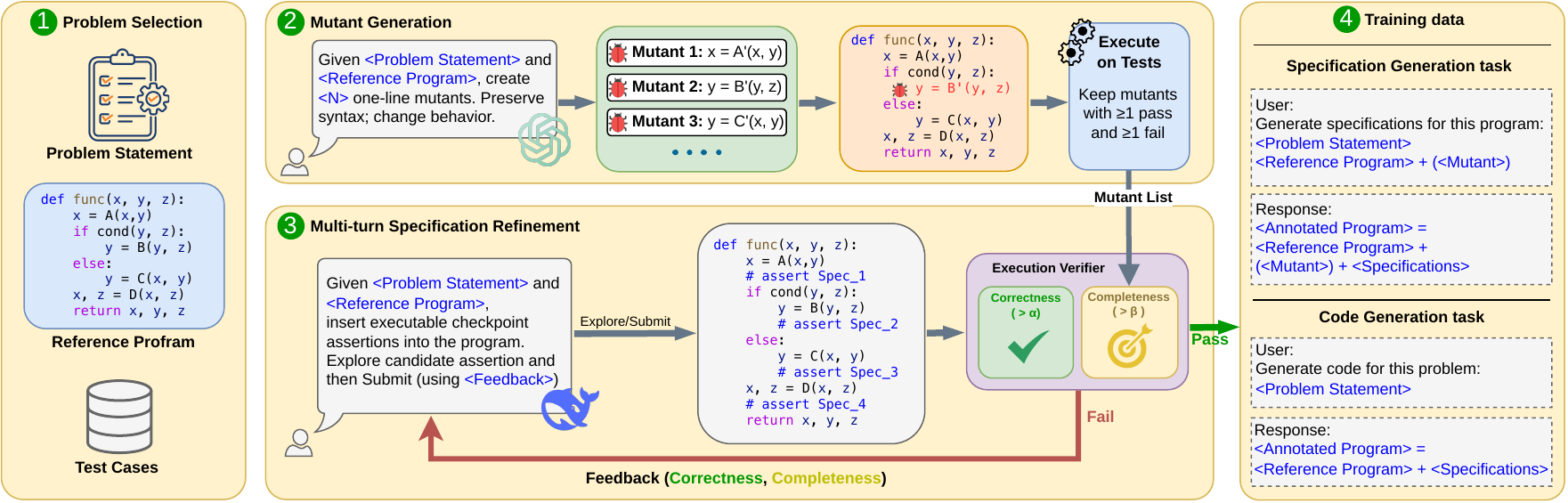}
\vspace{-6pt}
\caption{Pipeline for mutant-guided checkpoint specification data construction.}
\label{fig:speccoder}
\end{figure*}

We present \tool, a training framework for adapting CodeLLMs to generate executable checkpoint specifications. Given a programming problem and a program, \tool constructs supervision from assertions that are validated by execution rather than accepted solely from LLM output.


The pipeline consists of four stages. First, we select reliable reference programs from a LeetCode-style programming system by keeping only submissions that pass their available test suites. These executions provide the correct runtime states on which checkpoint assertions should hold. Second, we construct behavior-changing mutants from each reference program and retain only non-trivial mutants that pass at least one test and fail at least one test. These mutants provide controlled faulty executions, excluding equivalent variants and programs that are too broken to yield useful intermediate signals. 

Third, a teacher model generates checkpoint specifications through a multi-turn Explore--Submit process. At each turn, candidate assertions are instrumented and executed on the reference program and retained mutants. The verifier reports syntax errors, unsafe assertions, unreachable checkpoints, assertion failures, and mutant-rejection evidence. A submitted annotated program is accepted only when its assertions satisfy the required correctness and completeness thresholds. Finally, the accepted annotated programs are converted into supervised fine-tuning examples for two tasks: inserting checkpoint specifications into an existing program and generating a complete specification-bearing solution from the problem statement.
This ensures that the student model learns from checkpoint specifications that are reachable, executable, valid on reference executions, and challenged by behavior-changing mutants.

Fig.~\ref{fig:speccoder} illustrates the data construction pipeline, and
Algorithm~\ref{alg:speccoder-data} gives its procedural form. Starting from
problems, submissions, and test suites, we select validated reference programs,
generate behavior-changing mutants as controlled faulty executions, and use
multi-turn specification refinement to synthesize executable checkpoint
assertions. Candidate specifications are executed on both reference programs
and retained mutants, and only annotated programs satisfying the correctness
and completeness thresholds are kept for training.

\begin{algorithm}[t]
\caption{Mutant-guided checkpoint specification data construction}
\label{alg:speccoder-data}
\resizebox{0.85\linewidth}{!}{
\begin{minipage}{\linewidth}
\begin{algorithmic}[1]
\Require Problems \(\mathcal{P}\), submissions, test suites, teacher model \(M\),
thresholds \(\alpha,\beta\), maximum turns \(K\)
\Ensure Training data \(\mathcal{D}_{spec}, \mathcal{D}_{code}\)

\State \(\mathcal{D}_{spec}, \mathcal{D}_{code} \gets \emptyset\)

\ForAll{problem \(x_i \in \mathcal{P}\)}
    \State collect candidate program \(y_i\) and test suite \(T_i\)
    \If{\(|T_i| \leq 30\) or \(y_i\) does not pass all tests}
        \State \textbf{continue}
    \EndIf

    \State generate one-line mutants from \(y_i\)
    \State keep mutants with \(\geq 1\) pass and \(\geq 1\) fail
    \State deduplicate retained mutants to obtain \(\mathcal{M}_i\)

    \State \(F \gets \emptyset,\quad \tilde{y}_i \gets \emptyset\)

    \For{\(k=1,\ldots,K\)}
        \State \(a \gets M(x_i, y_i, F)\) \Comment{\textsc{Explore} or \textsc{Submit}}
        \State extract annotated program \(\hat{y}_i\) from \(a\)
        \State execute \(\hat{y}_i\) on \(T_i\) and \(\mathcal{M}_i\)
        \State compute \(\mathrm{Corr}_{ij}\) and \(\mathrm{Comp}_{ij}\) for all assertions

        \If{$a=\textsc{Submit}$ and $\textsc{Accept}(S_i,\alpha,\beta)$}
            \State \(\tilde{y}_i \gets \operatorname{Filter}(\hat{y}_i)\)
            \State \textbf{break}
        \EndIf

        \State update feedback \(F\) from execution results
    \EndFor

    \If{\(\tilde{y}_i = \emptyset\)}
        \State \textbf{continue}
    \EndIf

    \State add \((x_i, y_i) \rightarrow \tilde{y}_i\) to \(\mathcal{D}_{spec}\)
    \State add annotated mutant examples \((x_i,m)\rightarrow\tilde{m}\) to \(\mathcal{D}_{spec}\)
    \State add \(x_i \rightarrow \tilde{y}_i\) to \(\mathcal{D}_{code}\)
\EndFor

\State \Return \(\mathcal{D}_{spec}, \mathcal{D}_{code}\)
\end{algorithmic}
\textsc{Accept}$(S_i,\alpha,\beta)$ holds if every generated assertion is
reached on the reference code and satisfies
\(\mathrm{Corr}_{ij}\geq \alpha\), every assertion with an evaluable mutant set
satisfies \(\mathrm{Comp}_{ij}\geq \beta\), and at least one assertion has a
non-empty eligible mutant set.
\end{minipage}
}
\end{algorithm}


\subsubsection{Problem Selection and Reference Validation}

Given a programming problem \(x_i\), we collect a human-written submission
\(y_i\) and its test suite \(T_i\). To ensure that the constructed
specifications are grounded in executions, we keep a sample only when
two conditions are satisfied. First, the test suite must contain more than 30
tests, so that execution feedback covers more than a small number of examples.
Second, the submission must pass all available tests:
$
\forall t \in T_i,\quad \mathrm{Exec}(y_i,t)=\mathrm{Pass}.
$
This step gives us a trusted reference implementation. The executions of
\(y_i\) define the correct runtime states on which checkpoint specifications
should hold. The same reference program is also used as the base program for
constructing behavior-changing mutants. All train/validation/test splits are performed at the problem level before mutant generation and specification refinement. Thus, mutants and annotated programs derived from a problem cannot appear in training.

\subsubsection{Mutant Generation}

After validating the reference program, we construct mutants to provide negative
executions. For each executable statement in \(y_i\), we prompt an LLM to
generate plausible one-line buggy replacements. The prompt asks for natural
coding mistakes, such as using an incorrect condition, updating the wrong
variable, calling an incorrect helper function, or omitting a necessary update.
Each mutant must preserve syntax while changing program behavior.

We run every generated mutant on the same test suite \(T_i\). We keep only
non-trivial mutants that pass at least one test and fail at least one test:
$
0 <
\#\{t \in T_i \mid \mathrm{Exec}(m,t)=\mathrm{Pass}\}
<
|T_i|.
$
A mutant passing all tests is likely equivalent to the reference program
under the available tests, while a mutant that fails all tests is often too
broken to provide useful intermediate signals. We also deduplicate
mutants with the same pass/fail pattern over the test suite. The remaining
mutants form the retained mutant pool \(\mathcal{M}_i\). They serve as
controlled faulty executions: a useful checkpoint specification should hold on
the reference program but reject at least some faulty executions.

\subsubsection{Multi-turn Specification Refinement}

Inspired by the multi-turn refinement strategy of SpecMind~\cite{le-etal-2026-specmind},
we generate executable checkpoint specifications through a verification-guided
Explore--Submit process. For each task \(i\), the teacher model is given the
problem statement \(x_i\), a validated reference implementation \(y_i\), the
available functional tests \(T_i\), and a pool of retained behavior-changing
mutants \(M_i\). The goal is to insert inline checkpoint specifications into
\(y_i\) while leaving the executable implementation unchanged.


At each turn, the teacher either explores candidate checkpoints or submits an
annotated program for acceptance. Regardless of the turn type, the verifier
executes the candidate annotations on both the reference program and retained
mutants, producing feedback on reachability, correctness, and mutant rejection.
The difference is that an EXPLORE turn uses the feedback only to guide the next
revision, while a SUBMIT turn additionally triggers the acceptance test against
the predefined quality thresholds. The implementation code is required to remain byte-equivalent after removing assertion comments; turns
that modify executable code or contain unsafe assertion expressions are rejected
and regenerated.

We first extract all assertion comments from the annotated program and
instrument them as executable runtime checks. Each assertion is evaluated at its
original program location by wrapping its Boolean expression in a trace function.
This records both whether a checkpoint is reached and whether the
assertion holds every time it is reached. Assertions that contain syntactic
errors, refer to undefined state at runtime, or use unsafe operations
such as mutation, I/O, randomness, or dynamic evaluation are treated as invalid
feedback.

\paragraph{Correctness}
Correctness measures whether a generated checkpoint specification is valid on
the reference implementation. Let \(S_i=\{s_{i1},\ldots,s_{ik}\}\) be the
assertions extracted from a candidate annotated program. For assertion
\(s_{ij}\) and test \(t\in T_i\), let \(r_{ijt}\) be the number of times the
checkpoint is reached during the reference execution, and let \(p_{ijt}\) be
the number of those evaluations for which the assertion is true. The
per-assertion correctness score is
\[
\mathrm{Corr}_{ij} =
\frac{
\sum_{t\in T_i}
\mathbf{1}[r_{ijt}>0 \wedge p_{ijt}=r_{ijt}]
}{
\sum_{t\in T_i}
\mathbf{1}[r_{ijt}>0]
}.
\]
If the assertion is not reached by any reference execution, we
set \(\mathrm{Corr}_{ij}=0\). Thus, an assertion is correct only when every
reached evaluation on the reference implementation satisfies the predicate.
We aggregate at the test level rather than the raw evaluation level so that a
single long-running test with many repeated checkpoint evaluations does not
dominate the correctness score.


\paragraph{Completeness}
Completeness measures whether a correct checkpoint specification can expose
faulty intermediate states in retained mutants. 
A checkpoint should only be expected to reject faults that can influence the state observed at that checkpoint. Therefore, we restrict completeness evaluation to eligible mutants whose changed line occurs before the checkpoint and whose execution can reach the checkpoint.
We only evaluate a mutant
against a checkpoint when the mutant is eligible for that checkpoint: the
mutated source line must occur before the checkpoint, and the corresponding
original line and checkpoint must co-reach on at least one reference execution.
This prevents a checkpoint from being penalized for mutants that cannot
causally affect or reach it.

For assertion \(s_{ij}\), let \(E_{ij}\subseteq M_i\) be the eligible mutants
that reach the checkpoint in at least one mutant execution.~A mutant
\(m\in E_{ij}\) is rejected if \(s_{ij}\) fails on any reached~exec\-ution of that
mutant. The per-assertion completeness score is
\[
\mathrm{Comp}_{ij} =
\frac{
\#\{m\in E_{ij}: s_{ij}\text{ rejects }m\}
}{
\#E_{ij}
}.
\]
When \(E_{ij}\) is empty, no completeness claim is made for that assertion and
is excluded from the averaged completeness~score. 


The verifier reports per-assertion feedback after each turn, including
reachability, correctness, mutant rejection counts, skipped mutants, and
surviving mutants when available. The teacher uses this feedback to preserve
reliable assertions, repair false or unreachable ones, and strengthen assertions
that are correct but weak.

The acceptance test is applied at the assertion level rather than only to an
average score. This prevents a submitted program from being accepted because a
few strong checkpoints compensate for invalid, unreachable, or weak ones.
A submitted annotated program is accepted only if its assertions satisfy the
quality thresholds. That is, every generated assertion must be reached on
the reference implementation and satisfy
$
\mathrm{Corr}_{ij}\geq \alpha,
$
and every assertion with an evaluable mutant set must satisfy
$
\mathrm{Comp}_{ij}\geq \beta,
$
with at least one assertion receiving a non-vacuous completeness evaluation. In
our implementation, the default thresholds are \(\alpha=1.0\) and
\(\beta=0.9\). If no submitted program satisfies these conditions within the
maximum number of turns, the trajectory is marked unsuccessful and is not used
as accepted specification supervision.


\subsubsection{Training Data}

Accepted annotated programs are converted into two supervised fine-tuning tasks:
specification generation and code generation.

\paragraph{Specification generation}
This task teaches the model to insert executable checkpoint specifications into
an existing program. The input contains the problem statement and a program
\(p_i\), where \(p_i\) can be either the validated reference program or a
retained mutant:
\[
p_i \in \{y_i\} \cup \mathcal{M}_i .
\]
{\color{custom-blue}For the reference program, the target is the accepted annotated program
\(\tilde{y}_i\). For a mutant, we construct the target \(\tilde{m}\) by
inserting the accepted checkpoint specifications at the corresponding program
locations of the mutant, while preserving the mutant's executable code. The
target assertions describe the intended intermediate semantics of the task, not
the behavior of the buggy code. Thus, when \(\tilde{m}\) is executed,
some assertions may fail and expose the fault.}
The training example is the same program annotated with executable checkpoint
specifications:
\[
(x_i, p_i) \rightarrow \tilde{p}_i .
\]

\paragraph{Code generation}
This task teaches the model to generate a specification-bearing solution
directly from the problem statement. The input contains only the problem
statement, and the output is the validated reference solution annotated with
checkpoint specifications:
$
x_i \rightarrow \tilde{y}_i .
$
The final training data is the union of the two task datasets:
\[
\mathcal{D} = \mathcal{D}_{spec} \cup \mathcal{D}_{code}.
\]




\subsection{Model Training}
\label{sec:speccoder-training}

\tool is trained with supervised fine-tuning on the constructed instruction
data. The training set contains two types of examples: specification-generation
examples, where the model annotates a given program with executable checkpoint
assertions, and code-generation examples, where the model generates a complete
solution together with checkpoint assertions. The two tasks share the same goal:
the response should be a program with intermediate specifications.


All examples are formatted as instruction-response pairs. For specification
generation, the instruction contains the problem statement and an input program;
for code generation, the instruction contains only the problem statement. In
both cases, the response is the final annotated program. We do not train the
student model to reproduce the teacher's exploration turns or verifier
feedback. The multi-turn refinement process is used only to construct and filter
high-quality supervision targets.

With the training corpus
$
\mathcal{D} = \mathcal{D}_{spec} \cup \mathcal{D}_{code},
$
we optimize the standard autoregressive language-modeling objective over the
target response:
\[
\mathcal{L}(\theta)
=
-
\sum_{(u,o)\in \mathcal{D}}
\sum_{j=1}^{|o|}
\log p_{\theta}(o_j \mid u, o_{<j}),
\]
where \(u\) is the instruction input and \(o\) is the annotated-program
response.
At inference time, the trained model can be used in
the specification-generation setting to insert checkpoint assertions into a
given program.



\section{HumanExec Dataset}
\label{sec:humanexec}

We introduce HumanExec, a benchmark for evaluating whether models can generate and use executable specifications in realistic programming tasks. HumanExec is collected from recent Codeforces problems and is kept separate from the LeetCode-style data used to construct \tool's training corpus. Each task contains a problem statement, input/output specification, reference solution, human-submitted buggy solutions, and an executable test suite. The buggy programs come from real human submissions rather than synthetic mutation rules. Thus, the benchmark captures common programming mistakes in algorithm design, boundary handling, state updates, data-structure usage, and case analysis. This allows us to test if executable checkpoint specifications are useful beyond the synthetic mutants used during training-data construction.

To make the benchmark suitable for semantic verification, we filter out shallow failures. We remove submissions that do not compile, use malformed input/output handling, or fail only due to immediate runtime errors. We retain bugs that execute normally but produce incorrect outputs on some tests, since these cases require reasoning about the intended algorithm rather than merely fixing syntax or crashes. We also validate each reference solution against the available test suite and keep only tasks whose reference program passes all tests. For a buggy submission, we retain only non-trivial incorrect programs that pass at least one test and fail at least one test. This ensures that the benchmark focuses on semantic deviations rather than equivalent solutions or completely broken code.

HumanExec is aimed to measure both specification quality and downstream programming performance under a shared execution setting. It has a total of 150 tasks, each has 48--279 test cases with mean=79.53 and median=70. The number of mutant per task is 150, with min=3, max=188, mean=53.19, and median=44. The difficulty level is from 800-2500 (CodeForces standard), with mean=1348, and median=1300.


\noindent\textbf{Specification generation.}
Given a problem statement and a reference solution, the model must insert
executable checkpoint specifications at meaningful internal program points.
This subtask evaluates whether generated assertions are executable, valid on
correct executions, and discriminative against faulty executions from
human-submitted buggy programs.

\noindent\textbf{Program repair.}
Given a problem statement and a human-submitted buggy solution, the model must
produce a repaired solution that passes the test suite. This subtask tests
whether executable checkpoint specifications can provide useful local evidence
for identifying and correcting semantic errors.

\noindent\textbf{Program correctness checking.}
Given a problem statement and a candidate solution, the model must determine whether the solution is correct. This subtask evaluates whether models can use specification to justify correctness or expose a bug, instead of relying only on surface-level code plausibility.


\section{Empirical Evaluation}
\label{sec:evaluation}

We evaluate \tool from different perspectives: whether it generates higher-quality
executable checkpoint specifications for a given code, and whether these specifications improve
downstream programming tasks. Our evaluation is centered around the following
research questions:

\begin{itemize}
    \item \textbf{RQ1: Specification Generation.} Can \tool generate more
    correct and discriminative checkpoint specifications than prompted base
    models for a given code? 
    \item \textbf{RQ2: Program Correctness Checking.} Can {\tool}'s checkpoint specifications help models judge whether a candidate program is correct?
    \item \textbf{RQ3: Program Repair.} Do executable checkpoint
    specifications improve code repair of buggy programs?
\end{itemize}


\begin{table*}[t]
\centering
\caption{Specification-generation quality on HumanExec. \textbf{Abbreviations:} Corr. = correctness on correct executions; Comp. = completeness against behavior-changing mutants; \%Valid = executable assertion rate.}
\label{tab:rq1-spec-generation}
\scriptsize
\setlength{\tabcolsep}{3pt}
\renewcommand{\arraystretch}{0.8}
\vspace{-3pt}
\resizebox{0.9\textwidth}{!}{%
\begin{tabular}{@{}lcccccccccc@{}}
\toprule
\multirow{2}{*}{\textbf{Model}} 
& \multicolumn{6}{c}{\textbf{Inline Checkpoint Specs}}
& \multicolumn{4}{c}{\textbf{Postconditions}} \\
\cmidrule(lr){2-7} \cmidrule(l){8-11}
& \textbf{Corr.} 
& \textbf{$\Delta$ Corr.}
& \textbf{Comp.} 
& \textbf{$\Delta$ Comp.}
& \textbf{\%Valid} 
& \textbf{$\Delta$ Valid}
& \textbf{Corr.} 
& \textbf{$\Delta$ Corr.}
& \textbf{Comp.} 
& \textbf{$\Delta$ Comp.} \\
\midrule

Qwen2.5-Coder-7B
& 0.2794 & \basedelta
& 0.1707 & \basedelta
& 0.8272 & \basedelta
& 0.5358 & \basedelta
& 0.3161 & \basedelta \\
\textbf{\emph{+ SpecCoder}}
& \textbf{0.4353} & \textbf{\posdelta{55.8}}
& \textbf{0.7820} & \textbf{\posdelta{358.1}}
& \textbf{0.9028} & \textbf{\posdelta{9.1}}
& \textbf{0.5389} & \textbf{\posdelta{0.6}}
& \textbf{0.6981} & \textbf{\posdelta{120.8}}\\
\midrule

Qwen2.5-Coder-14B
& 0.5227 & \basedelta
& 0.1989 & \basedelta
& 0.7900 & \basedelta
& 0.5808 & \basedelta
& 0.1901 & \basedelta \\
\textbf{\emph{+ SpecCoder}}
& \textbf{0.6468} & \textbf{\posdelta{23.7}}
& \textbf{0.7713} & \textbf{\posdelta{287.8}}
& \textbf{1.0000} & \textbf{\posdelta{26.6}}
& \textbf{0.6477} & \textbf{\posdelta{11.5}}
& \textbf{0.6643} & \textbf{\posdelta{249.4}} \\
\midrule

Qwen2.5-Coder-32B
& 0.6313 & \basedelta
& 0.2640 & \basedelta
& 0.8989 & \basedelta
& 0.7697 & \basedelta
& 0.2394 & \basedelta \\
\textbf{\emph{+ SpecCoder}}
& \textbf{0.8098} & \textbf{\posdelta{28.3}}
& \textbf{0.6843} & \textbf{\posdelta{159.2}}
& \textbf{0.9973} & \textbf{\posdelta{10.9}}
& \textbf{0.9047} & \textbf{\posdelta{17.5}}
& \textbf{0.4561} & \textbf{\posdelta{90.5}} \\
\bottomrule
\end{tabular}%
}
\end{table*}

\subsubsection{Datasets}
We use HumanExec as the primary benchmark for all three RQs. 
For RQ1, we evaluate specification generation on reference solutions. Generated assertions are executed on correct reference executions to measure validity and correctness, and are evaluated against faulty executions to measure discriminativeness. These faulty executions are obtained from human-submitted buggy solutions in HumanExec; if evaluation-only mutants are additionally used, they are~gene\-rated only after the train/test split and are never used for training. For RQ2, we construct candidate programs from both reference solutions and human-submitted buggy solutions and ask models to predict if each candidate is correct. For RQ3, we use human-submitted buggy solutions as repair inputs and evaluate generated repairs against the HumanExec test~suites.


\subsubsection{Models}
We evaluate representative LLMs from the Qwen2.5-Coder families: Qwen2.5-Coder-7B, Qwen2.5-Coder-14B, Qwen2.5-Coder-32B. For each model, we compare the
base instruction-tuned model with its \tool-tuned counterpart.
All methods use the same decoding settings and evaluation test suites.

\subsection{RQ1: Specification Generation}
\label{sec:eval-rq1-specgen}

\subsubsection{Baselines}
We compare \tool against zero-shot specification generation, where the base instruction-tuned model directly annotates a given program with checkpoint specifications. This baseline tests whether prompting alone is sufficient to produce high-quality executable specifications from the same backbone model.

\subsubsection{Metrics}
We evaluate generated specifications using execution-based metrics. \emph{Validity} measures the fraction of generated assertions that can be safely instrumented and executed, excluding assertions with syntax errors, out-of-scope variables, or unsafe operations. \emph{Correctness} measures whether a reachable assertion holds on correct executions of the reference solution. \emph{Completeness} measures, in an operational mutant-based sense, whether a correct assertion rejects behavior-changing faulty executions that reach the same checkpoint. {\color{custom-blue}Completeness is measured using evaluation-only behavior-changing mutants constructed from HumanExec reference solutions after the train/test split}. These mutants are used only for evaluation and are not included in {\tool}'s training data. Note that completeness captures discriminative power against the available faulty executions rather than logical completeness over all possible bugs. We report these metrics separately for inline checkpoint specifications and final postconditions.

\begin{figure}
    \centering
    \includegraphics[width=1\linewidth]{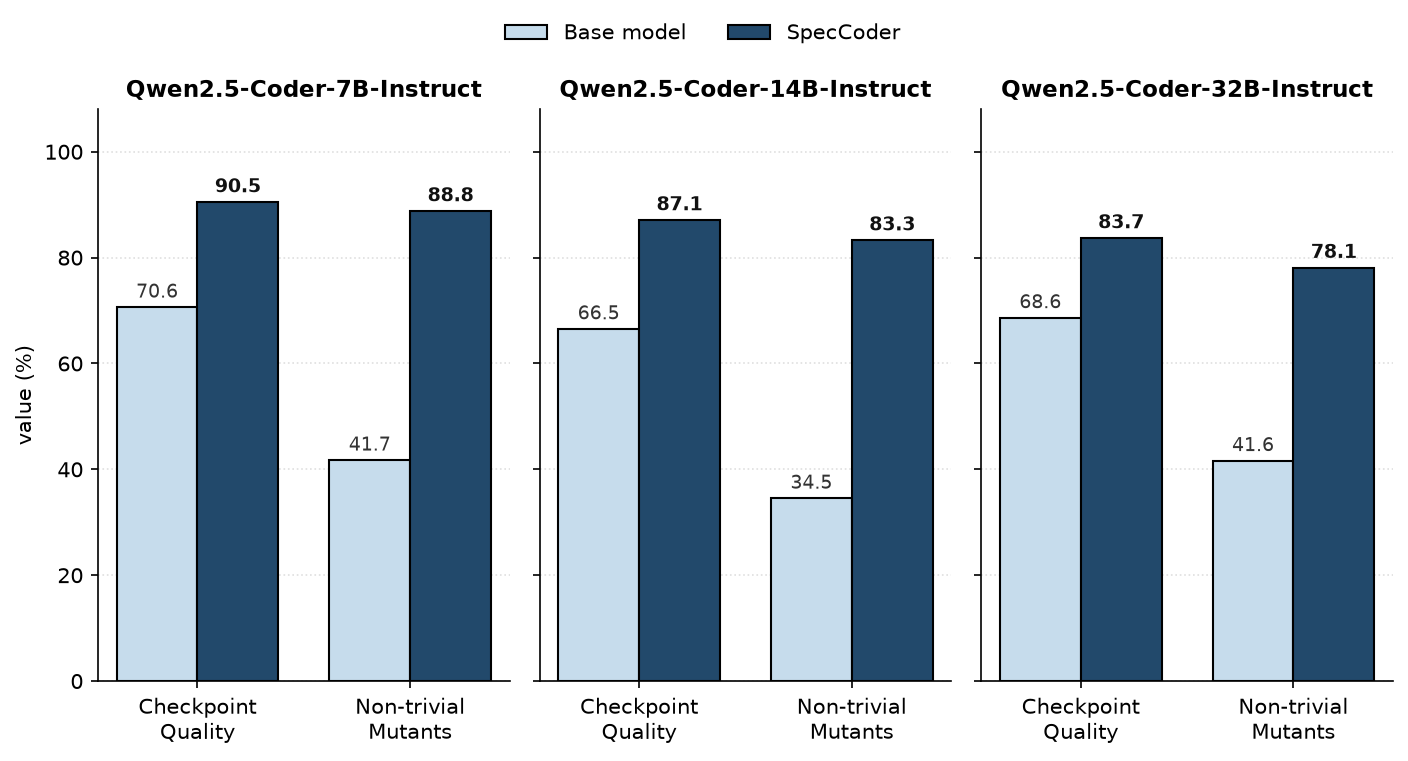}
    \vspace{-20pt}
    \caption{Quality of generated checkpoint assertions.}
    \label{fig:checkpoint-quality}
\end{figure}


\subsubsection{Results}
Table~\ref{tab:rq1-spec-generation} summarizes the specification-generation
results on HumanExec. Across the zero-shot base models, 
larger base models improve correctness and completeness, although validity is
not strictly monotonic.~Yet, their specifications remain limited. Qwen2.5-Coder-7B achieves only 0.2794 correctness and 0.1707 completeness, while Qwen2.5-Coder-14B improves
correctness to 0.5227 but still obtains only 0.1989 completeness. The strongest
base model, Qwen2.5-Coder-32B, further improves to 0.6313 correctness and
0.2640 completeness, yet its generated specifications are still far from
reliably discriminative. These results indicate that scaling the base model
helps, but prompting alone is insufficient for generating specifications that
are both correct and useful for rejecting buggy executions.

\tool substantially improves inline checkpoint specification quality. With
Qwen2.5-Coder-14B, \tool increases correctness from 0.5227 to 0.6468 and
completeness from 0.1989 to 0.7713, while reaching 1.0000 executable assertion
validity. With Qwen2.5-Coder-32B, \tool reaches 0.8098 correctness and 0.6843
completeness, compared with 0.6313 and 0.2640 for the corresponding base
zero-shot model. 
%
Across the three tuned backbones, \tool improves average inline correctness from 0.4778 to 0.6306, average inline completeness from 0.2112 to 0.7459, and average validity from 0.8387 to 0.9667.
These gains show that verification-guided fine-tuning teaches the model to generate checkpoint assertions that are not only syntactically executable, but also substantially
more discriminative against behavior-changing mutants. The specifications from {\tool} range 
from value constraints, ordering structure, loop progress, algorithm, data shape, and others.

Figure~\ref{fig:checkpoint-quality} further explains the source of these gains. We measure \textbf{Checkpoint Quality} as the fraction of assertions reached by at least one test case, and \textbf{Non-triviality} as the fraction of assertions that reject at least one mutant. 
These metrics are useful because a model could generate assertions in locations where no faulty execution reaches or can be affected by them. Such assertions may be correct, but they do not provide evidence for discriminating correct behavior from faulty behavior.
Across all three backbones, \tool consistently improves both metrics: average checkpoint quality increases from 68.6\% to 87.1\%, while the non-trivial assertion rate increases from 39.3\% to 83.4\%. This shows that \tool does not merely generate more executable assertions; it generates assertions that are more often exercised during execution and more likely to expose faulty intermediate states.

The remaining results follow the same trend. For Qwen2.5-Coder-7B, \tool
achieves 0.4353 correctness, 0.7820 completeness, and 0.9028 validity,
improving over the corresponding zero-shot base model by 55.8\%, 358.1\%, and
9.1\%, respectively. For postcondition specifications, \tool achieves 0.6971
correctness and 0.6062 completeness on average, compared with 0.6288
correctness and 0.2485 completeness for zero-shot prompting. The postcondition
results show that \tool also improves final-state specifications, but the
largest gains occur for inline checkpoint completeness. This suggests that
verification-guided training is especially beneficial for the harder task of
generating intermediate assertions that expose faulty internal states. Overall,
the results indicate that executable checkpoint specifications require
verification-guided training rather than direct prompting alone.


\subsubsection{Does SpecCoder Internalize Checkpoint Specifications?}

We further measure whether \tool internalizes executable checkpoint
specifications using 30 held-out validation problems from the dataset
construction process. Each problem contains verified annotated specifications
produced by our verification-guided refinement pipeline. For each assertion, we
teacher-force the same annotated program under the base model and the
corresponding \tool model, and compute perplexity only over assertion tokens.
Since both models score the same target assertions, lower perplexity indicates
that the model assigns higher likelihood to the same specification.

\begin{figure}
    \centering
    \vspace{-18pt}
    \includegraphics[width=1\linewidth]{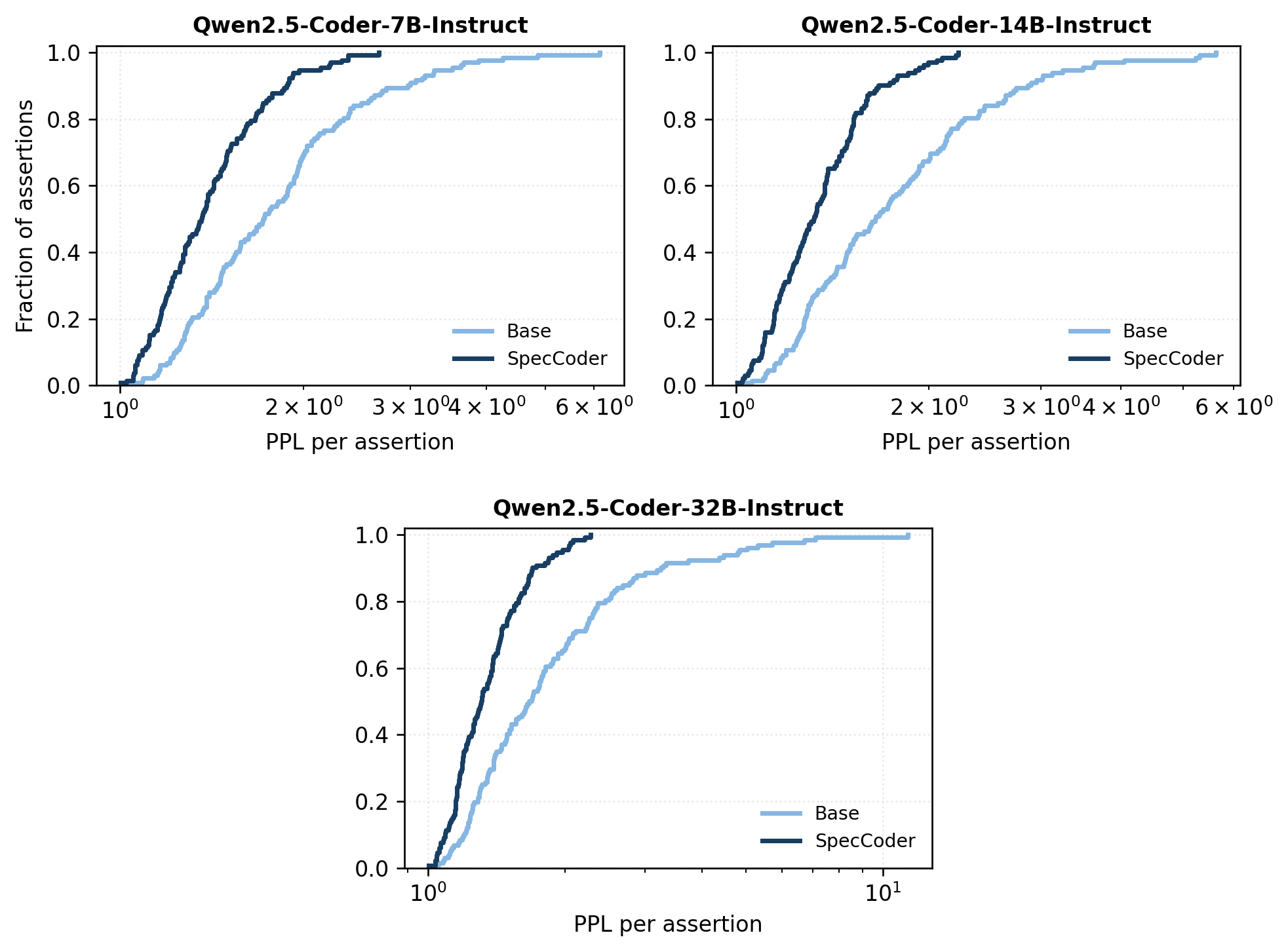}
    \vspace{-18pt}
    \caption{ECDF of per-assertion perplexity on held-out inline checkpoint
    specifications. Each panel compares a base Qwen2.5-Coder model with its
    \tool-tuned counterpart on the same assertion tokens.}
    \label{fig:assertion-ppl-ecdf}
\end{figure}

Fig.~\ref{fig:assertion-ppl-ecdf} shows that \tool shifts the per-assertion
perplexity distribution left across all three backbones, indicating higher
confidence on the same executable assertions. Together with
Table~\ref{tab:rq1-spec-generation}, this suggests that it learns both the
syntax and placement patterns of checkpoint specifications, rather than relying
only on prompting-time instruction following.


\subsubsection{Qualitative Analysis}
\begin{figure*}[t]
\centering
\includegraphics[width=\linewidth]{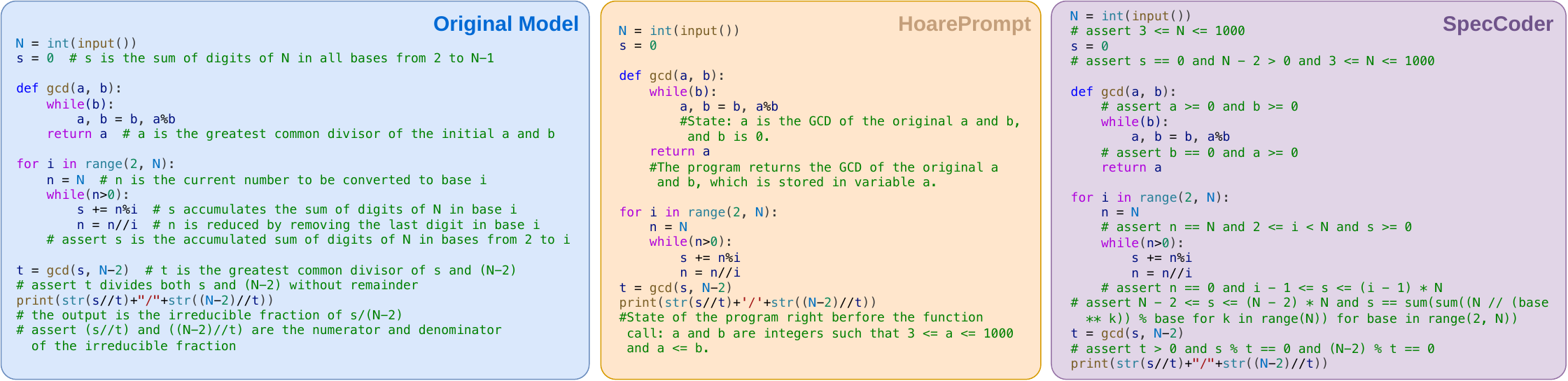}
\vspace{-18pt}
\caption{Qualitative comparison of specification generation}
\label{fig:qualitative_example}
\end{figure*}

Fig.~\ref{fig:qualitative_example} shows a representative example comparing zero-shot specification generation with \tool. The original model does not generate specifications in the required executable format; instead, it mainly inserts explanatory comments, such as describing that \texttt{s} accumulates digit sums or that \texttt{t} is a greatest common divisor. {\em Even comments prefixed with ``assert'' are still natural-language descriptions} rather than valid Boolean predicates. As a result, they cannot be directly instrumented or checked during execution. In contrast, \tool generates executable checkpoint assertions over in-scope variables, such as \code{s == 0 and N - 2 > 0}, \code{n == N and 2 <= i < N}, and \code{t > 0 and s \% t == 0 and (N-2) \% t == 0}. These assertions are placed at meaningful internal program points and can be evaluated on runtime states, illustrating why \tool improves both assertion validity and discriminative completeness.

\subsection{RQ2: Program Correctness Checking}
\label{sec:eval-rq2-correctness}

\begin{table*}[t]
\centering
\caption{Program correctness-checking performance on \textsc{HumanExec}. 
\textbf{Abbreviations:} MCC = Matthews Correlation Coefficient; 
$\Delta$MCC = relative MCC improvement over Vanilla; 
BA = Balanced Accuracy; 
TPR = True Positive Rate; 
FNR = False Negative Rate; 
FPR = False Positive Rate; 
TNR = True Negative Rate.}
\label{tab:rq2-correctness-checking}
\footnotesize
\setlength{\tabcolsep}{4.0pt}
\renewcommand{\arraystretch}{1.08}
\resizebox{0.9\textwidth}{!}{%
\begin{tabular}{@{}
    >{\raggedright\arraybackslash}p{2.4cm}
    >{\raggedright\arraybackslash}p{2cm}
    *{10}{c}
@{}}
\toprule
\multirow{2}{*}{\textbf{Model}}
& \multirow{2}{*}{\textbf{Classifier}}
& \multicolumn{7}{c}{\textbf{Correctness-Checking Metrics}}
& \multicolumn{3}{c}{\textbf{Tokens (Millions)}} \\
\cmidrule(lr){3-9} \cmidrule(l){10-12}
&
& \textbf{MCC$\uparrow$} 
& \textbf{$\Delta$MCC$\uparrow$} 
& \textbf{BA$\uparrow$} 
& \textbf{TNR$\uparrow$}
& \textbf{TPR$\uparrow$} 
& \textbf{FNR$\downarrow$} 
& \textbf{FPR$\downarrow$} 
& \textbf{Input} 
& \textbf{Output} 
& \textbf{Total} \\
\midrule

\multirow{3}{=}{Qwen2.5-Coder-7B}
& Vanilla      & 0.236 & \basedelta        & 0.600 & 0.733 & 0.167 & 0.333 & 0.267 & 0.38 & 0.21 & 0.59 \\
& CoT          & 0.211 & \textbf{\negdelta{-10.71}} & 0.600 & 0.773 & 0.220 & 0.280 & 0.227 & 0.45 & 0.31 & 0.76 \\
& HoarePrompt  & 0.310 & \textbf{\posdelta{30.94}}  & 0.639 & 0.713 & 0.427 & 0.073 & 0.287 & 59.30 & 8.93 & 68.23 \\
\textbf{+ \emph{SpecCoder}} 
& AssertChecker
               & \textbf{0.307} & \textbf{\posdelta{29.75}} & 0.653 & 0.820 & 0.333 & 0.167 & 0.180 & 0.82 & 0.48 & 1.30 \\
\midrule

\multirow{3}{=}{Qwen2.5-Coder-14B}
& Vanilla      & 0.283 & \basedelta        & 0.633 & 0.900 & 0.233 & 0.267 & 0.100 & 0.38 & 0.22 & 0.60 \\
& CoT          & 0.229 & \textbf{\negdelta{-19.14}} & 0.613 & 0.840 & 0.273 & 0.227 & 0.160 & 0.45 & 0.38 & 0.83 \\
& HoarePrompt  & 0.366 & \textbf{\posdelta{29.56}}  & 0.680 & 0.887 & 0.293 & 0.207 & 0.113 & 53.01 & 10.47 & 63.58 \\
\textbf{+ \emph{SpecCoder}} 
& AssertChecker
               & \textbf{0.384} & \textbf{\posdelta{35.69}} & 0.680 & 0.887 & 0.253 & 0.247 & 0.113 & 0.91 & 0.34 & 1.25 \\
\midrule

\multirow{3}{=}{Qwen2.5-Coder-32B}
& Vanilla      & 0.394 & \basedelta        & 0.693 & 0.893 & 0.300 & 0.200 & 0.107 & 0.38 & 0.08 & 0.46 \\
& CoT          & 0.248 & \textbf{\negdelta{-36.97}} & 0.620 & 0.873 & 0.247 & 0.253 & 0.127 & 0.45 & 0.09 & 0.18 \\
& HoarePrompt  & 0.455 & \textbf{\posdelta{15.69}}  & 0.727 & 0.840 & 0.387 & 0.113 & 0.160 & 48.82 & 12.09  & 60.91 \\
\textbf{+ \emph{SpecCoder}} 
& AssertChecker
               & \textbf{0.493} & \textbf{\posdelta{25.15}} & 0.733 & 0.947 & 0.287 & 0.213 & 0.053 & 0.87 & 0.49 & 1.36 \\
\bottomrule
\end{tabular}%
}
\end{table*}

RQ2 evaluates whether checkpoint specifications help a model judge whether a
candidate program is correct. This task is different from repair: the model
must produce a correctness verdict and justification rather than a patch.

\subsubsection{Baselines}
We compare direct correctness checking from the problem statement and code
against chain-of-thought prompting, HoarePrompt-style natural-language
specification reasoning, and \tool{} with executable assertion evidence. The
\tool{} classifier receives checkpoint specifications and, when applicable,
violated-specification evidence.

\subsubsection{Metrics}
We report Matthews Correlation Coefficient (MCC), balanced accuracy, true
negative rate, true positive rate, false negative rate, and false positive
rate. MCC and balanced accuracy summarize overall classification quality under
class imbalance, while FPR and FNR distinguish the risk of incorrectly
accepting a buggy program from the risk of incorrectly rejecting a correct
program.

\subsubsection{Results}
Table~\ref{tab:rq2-correctness-checking} reports correctness-checking results
on \textsc{HumanExec}. Overall, executable checkpoint evidence improves correctness checking across model sizes. Compared with Vanilla, \tool{} improves MCC from 0.236 to 0.307 for Qwen2.5-Coder-7B, from 0.283 to 0.384 for Qwen2.5-Coder-14B, and from 0.394 to 0.493 for Qwen2.5-Coder-32B. Balanced accuracy follows the same trend: \tool{} achieves 0.653, 0.680, and 0.733 across the three backbones, matching or exceeding the strongest baseline in each model group. In contrast, simple chain-of-thought prompting does not reliably improve correctness checking and even reduces MCC for all three backbones. This suggests that free-form reasoning alone is insufficient, while executable specification evidence gives the checker a more reliable signal.


\subsubsection{Qualitative Analysis}
Figure~\ref{fig:qualitative_example} also illustrates why executable checkpoint specifications provide stronger evidence for correctness checking than natural-language state reasoning. HoarePrompt generates readable state descriptions, such as explaining that \texttt{a} is the GCD of the original arguments or describing the state before a function call, but these descriptions are not executable and cannot be objectively checked against a candidate program. Some conditions also refer to implicit concepts such as ``original'' variable values, making them difficult to bind to concrete runtime states. By contrast, \tool produces executable assertions that define concrete obligations over program variables, including initialization constraints, loop-state properties, accumulated digit-sum relations, and divisibility checks after \texttt{gcd}. When a candidate program violates such an assertion, the checker receives localized, machine-checkable evidence of the semantic deviation, explaining the improved correctness-checking performance of \tool.
\subsection{RQ3: Program Repair}
\label{sec:eval-rq3-repair}

\begin{table}[t]
\centering
\caption{Program repair performance on \textsc{HumanExec}. \textbf{Abbreviations:} Pass@\(k\) reports whether at least one of the top-\(k\) generated repairs passes the full test suite. APR reports the mean pass rate over generated repairs.}
\label{tab:rq3-program-repair}
\scriptsize
\setlength{\tabcolsep}{3.0pt}
\renewcommand{\arraystretch}{1.03}
\resizebox{0.9\columnwidth}{!}{%
\begin{tabular}{@{}llccc@{}}
\toprule
\textbf{Model}
& \textbf{Method}
& \textbf{Pass@1}
& \textbf{Pass@3}
& \textbf{APR} \\
\midrule

\multirow{3}{*}{Qwen2.5-Coder-7B}
& Vanilla      & 0.089 & 0.157 & 0.447 \\
& HoarePrompt  & 0.099 & 0.111 & 0.511 \\
& SpecRover    & 0.132 & 0.151 & \textbf{0.580} \\
\textbf{+ \emph{SpecCoder}}
& \textbf{AssertRepair}
               & \textbf{0.140} & \textbf{0.190} & 0.514 \\
\midrule

\multirow{3}{*}{Qwen2.5-Coder-14B}
& Vanilla      & 0.237 & 0.269 & 0.578 \\
& HoarePrompt  & 0.211 & 0.289 & 0.534 \\
& SpecRover    & 0.235 & 0.261 & 0.556 \\
\textbf{+ \emph{SpecCoder}}
& \textbf{AssertRepair}
               & \textbf{0.241} & \textbf{0.301} & \textbf{0.598} \\
\midrule

\multirow{3}{*}{Qwen2.5-Coder-32B}
& Vanilla      & 0.261 & 0.301 & 0.615 \\
& HoarePrompt  & 0.272 & \textbf{0.358} & 0.587 \\
& SpecRover    & 0.279 & 0.338 & 0.610 \\
\textbf{+ \emph{SpecCoder}}
& \textbf{AssertRepair}
               & \textbf{0.283} & 0.333 & \textbf{0.622} \\
\bottomrule
\end{tabular}%
}
\end{table}
RQ3 evaluates whether executable checkpoint specifications help models repair
human-written bugs. Given a buggy program, each method generates one or more
candidate repairs under the AgentCoder backbone. We execute each candidate
against the full test suite and measure whether the generated patch fixes the
program.

\subsubsection{Baselines}
We introduce AssertRepair, an APR method that leverages assertion signals for program repair. Specifically, given a failing test input, AssertRepair executes the program and captures the assertion states, including violated checkpoint specifications. The resulting diagnostic evidence is then provided to the LLM to guide repair generation.

Besides the vanilla baseline, we compare AssertRepair with state-of-the-art repair baselines based on natural-language specification signals. We adopt HoarePrompt (V), which uses programs annotated with HoarePrompt-generated specifications for repair, and SpecRover (V), a competitive-programming variant of SpecRover. In contrast to these natural-language-specification baselines, AssertRepair uses executable checkpoint specifications and violated-assertion evidence. To ensure a fair comparison, all methods are allowed only a single repair attempt per turn.

\subsubsection{Metrics}
We report \emph{Pass@k}, where \(k \in \{1,3\}\), measuring whether at least
one of the top-\(k\) generated repairs passes all tests. We also report Average Pass Rate (APR).

\subsubsection{Results}
Table~\ref{tab:rq3-program-repair} reports program repair results on
\textsc{HumanExec}. 
Overall, executable checkpoint evidence improves repair effectiveness, especially for the top-ranked generated repairs. \tool achieves the best Pass@1 for all three backbones: 0.140 for Qwen2.5-Coder-7B, 0.241 for Qwen2.5-Coder-14B, and 0.283 for Qwen2.5-Coder-32B.

For Qwen2.5-Coder-14B and Qwen2.5-Coder-32B, \tool achieves the highest APR, improving over Vanilla from 0.578 to 0.598 and from 0.615 to 0.622, respectively. These gains suggest that executable checkpoint specifications help the repair agent produce a larger fraction of passing patches, not only a better top candidate.

Averaged across the three backbones, \tool obtains an APR of 0.578, compared with 0.547 for Vanilla and 0.544 for HoarePrompt. Its average APR is close to SpecRover's 0.582, while \tool achieves stronger Pass@1 and Pass@3 on average. The results therefore suggest that executable checkpoint evidence is most effective at guiding the repair process toward correct high-ranked patches, while its effect on the entire candidate distribution depends on the strength of the underlying model.

Overall, RQ3 shows that executable checkpoint specifications provide useful localized evidence for program repair. 
This evidence improves top-ranked repair success across all model sizes and improves average repair quality for the stronger 14B and 32B backbones.

\section{Related Work}
\label{sec:related}

\paragraph{Traditional specification inference}
Dynamic approaches infer likely properties by
observing executions~\cite{ammons-popl02,BeschastnikhBSSE2011,daikon99}, but
their results depend heavily on the coverage and diversity of available tests.
Static analyses~\cite{10.1145/1273463.1273487,engler-sosp01,kremenek06,ramanathan-pldi07,yiwei-icse11}
and abstract interpretation~\cite{10.1145/2384616.2384633} often produce conservative or imprecise
specifications due to over-approximation and false positives. Other approaches
extract specifications from API documentation or code comments~\cite{10.5555/2337223.2337319,10.1145/1294261.1294276,10.1145/1985793.1985796,10.1109/ICST.2012.106,10.1145/3213846.3213872,10.1109/ASE.2009.94,7985647}.
Data-mining techniques infer common API usage patterns from large
codebases~\cite{prminer-fse05,dynamine05,mapo09}, including call pairs,
sequences, and finite-state models~\cite{zeller07,williams-tse05,taoxie-ase09,fse09,mike-ase09,zeller-ase09}.
These methods are useful for API usage constraints, but they rarely infer
 executable postconditions~\cite{ramanathan-pldi07}.

\paragraph{Machine-learning and LLM-based specification generation}
Recent machine-learning and LLM-based techniques have expanded automated
specification and oracle generation. Some methods synthesize test oracles or
unit tests~\cite{dinella2022toga,mastropaolo2023usingtransfer,10.1145/3524481.3527220,lahiri2023interactivecodegenerationtestdriven,tufano2021unittestcasegeneration},
while others improve test generation and coverage~\cite{lemieux2023codamosa,coverup2025}.
For example, AthenaTest~\cite{tufano2021unittestcasegeneration} generates unit
test inputs and oracles from the focal method implementation, TOGA~\cite{dinella2022toga}
focuses on oracle generation, and TiCoder~\cite{lahiri2023interactivecodegenerationtestdriven}
uses LLMs to generate inputs/outputs from textual intent. 


Other work directly targets specifications. NL2Postcond~\cite{nl2postcond}
uses LLMs to infer postconditions from code and textual descriptions, and
EvoSpex~\cite{10.1109/ICSE43902.2021.00112} applies evolutionary learning to
infer functional input--output relationships. Recent efforts also explore
property-based specifications: Vikram \emph{et al.}~\cite{vikram2024largelanguagemodelswrite}
use LLMs to generate property-based tests, while Speculyzer~\cite{DBLP:journals/corr/abs-2210-00848}
enumerates likely properties and candidate inputs to guide code generation.
Beyond input--output properties, learning-based methods have been used to infer
program invariants~\cite{10.1145/2837614.2837664,Laich2020Guiding,10.1145/3385412.3385986,pmlr-v202-pei23a}.
In contrast, \tool targets executable checkpoint
specifications: assertions placed at meaningful internal program points and
selected using execution feedback from correct programs and behavior-changing
mutants.

\paragraph{Specification-guided reasoning and repair.}
LLMs have been used to produce specification-like reasoning for
correctness checking and program repair. HoarePrompt~\cite{hoareprompt} uses
natural-language state conditions to support structural reasoning about program
correctness. SpecRover~\cite{ruan2025specrover} extracts natural-language code
intent to guide agentic program repair, vet patches, and provide supporting
evidence. These approaches show that specification-like signals can improve
LLM-based reasoning, but their intermediate signals are primarily expressed in
natural language and are not directly executable assertions. \tool differs by
training CodeLLMs to generate executable intermediate specifications that can
be checked on concrete executions and used as verifiable evidence for
specification generation, consistency checking, and repair.




\section{Conclusion}
\label{sec:conclusion}

We presented \tool, a verification-guided framework for training CodeLLMs to
generate executable checkpoint specifications. By learning from validated
reference programs, behavior-changing mutants, and refinement traces, \tool
produces assertions that capture intermediate program states and can be checked
during execution. Experiments on \textsc{HumanExec} show that these checkpoints
improve specification quality and provide useful evidence for correctness
checking and program repair, suggesting that executable intermediate
specifications are a promising path toward more verifiable and repairable
LLM-generated programs.

\section{Data Availability Statement} All data and code are available at~\cite{speccoder-website}.



\bibliographystyle{IEEEtran}

\bibliography{references,refs-specmining}

@article{nl2postcond,
author = {Endres, Madeline and Fakhoury, Sarah and Chakraborty, Saikat and Lahiri, Shuvendu K.},
title = {Can Large Language Models Transform Natural Language Intent into Formal Method Postconditions?},
year = {2024},
issue_date = {July 2024},
publisher = {Association for Computing Machinery},
address = {New York, NY, USA},
volume = {1},
number = {FSE},
url = {https://doi.org/10.1145/3660791},
doi = {10.1145/3660791},
journal = {Proc. ACM Softw. Eng.},
month = jul,
articleno = {84},
numpages = {24}
}

@inproceedings{10.1145/1273463.1273487,
author = {Shoham, Sharon and Yahav, Eran and Fink, Stephen and Pistoia, Marco},
title = {Static specification mining using automata-based abstractions},
year = {2007},
isbn = {9781595937346},
publisher = {Association for Computing Machinery},
address = {New York, NY, USA},
url = {https://doi.org/10.1145/1273463.1273487},
doi = {10.1145/1273463.1273487},
booktitle = {Proceedings of the 2007 International Symposium on Software Testing and Analysis},
pages = {174–184},
numpages = {11},
location = {London, United Kingdom},
series = {ISSTA '07}
}

@inproceedings{10.1145/2384616.2384633,
author = {Cousot, Patrick M. and Cousot, Radhia and Logozzo, Francesco and Barnett, Michael},
title = {An abstract interpretation framework for refactoring with application to extract methods with contracts},
year = {2012},
isbn = {9781450315616},
publisher = {Association for Computing Machinery},
address = {New York, NY, USA},
url = {https://doi.org/10.1145/2384616.2384633},
doi = {10.1145/2384616.2384633},
booktitle = {Proceedings of the ACM International Conference on Object Oriented Programming Systems Languages and Applications},
pages = {213–232},
numpages = {20},
location = {Tucson, Arizona, USA},
series = {OOPSLA '12}
}

@inproceedings{10.5555/2337223.2337319,
author = {Pandita, Rahul and Xiao, Xusheng and Zhong, Hao and Xie, Tao and Oney, Stephen and Paradkar, Amit},
title = {Inferring method specifications from natural language API descriptions},
year = {2012},
isbn = {9781467310673},
publisher = {IEEE Press},
booktitle = {Proceedings of the 34th International Conference on Software Engineering},
pages = {815–825},
numpages = {11},
location = {Zurich, Switzerland},
series = {ICSE '12}
}

@inproceedings{10.1145/1294261.1294276,
author = {Tan, Lin and Yuan, Ding and Krishna, Gopal and Zhou, Yuanyuan},
title = {/*icomment: bugs or bad comments?*/},
year = {2007},
isbn = {9781595935915},
publisher = {Association for Computing Machinery},
address = {New York, NY, USA},
url = {https://doi.org/10.1145/1294261.1294276},
doi = {10.1145/1294261.1294276},
booktitle = {Proceedings of Twenty-First ACM SIGOPS Symposium on Operating Systems Principles},
pages = {145–158},
numpages = {14},
location = {Stevenson, Washington, USA},
series = {SOSP '07}
}

@inproceedings{10.1145/1985793.1985796,
author = {Tan, Lin and Zhou, Yuanyuan and Padioleau, Yoann},
title = {aComment: mining annotations from comments and code to detect interrupt related concurrency bugs},
year = {2011},
isbn = {9781450304450},
publisher = {Association for Computing Machinery},
address = {New York, NY, USA},
url = {https://doi.org/10.1145/1985793.1985796},
doi = {10.1145/1985793.1985796},
booktitle = {Proceedings of the 33rd International Conference on Software Engineering},
pages = {11–20},
numpages = {10},
location = {Waikiki, Honolulu, HI, USA},
series = {ICSE '11}
}

@inproceedings{10.1109/ICST.2012.106,
author = {Tan, Shin Hwei and Marinov, Darko and Tan, Lin and Leavens, Gary T.},
title = {@tComment: Testing Javadoc Comments to Detect Comment-Code Inconsistencies},
year = {2012},
isbn = {9780769546704},
publisher = {IEEE Computer Society},
address = {USA},
url = {https://doi.org/10.1109/ICST.2012.106},
doi = {10.1109/ICST.2012.106},
booktitle = {Proceedings of the 2012 IEEE Fifth International Conference on Software Testing, Verification and Validation},
pages = {260–269},
numpages = {10},
series = {ICST '12}
}

@inproceedings{10.1145/3213846.3213872,
author = {Blasi, Arianna and Goffi, Alberto and Kuznetsov, Konstantin and Gorla, Alessandra and Ernst, Michael D. and Pezz\`{e}, Mauro and Castellanos, Sergio Delgado},
title = {Translating code comments to procedure specifications},
year = {2018},
isbn = {9781450356992},
publisher = {Association for Computing Machinery},
address = {New York, NY, USA},
url = {https://doi.org/10.1145/3213846.3213872},
doi = {10.1145/3213846.3213872},
booktitle = {Proceedings of the 27th ACM SIGSOFT International Symposium on Software Testing and Analysis},
pages = {242–253},
numpages = {12},
location = {Amsterdam, Netherlands},
series = {ISSTA 2018}
}

@inproceedings{10.1109/ASE.2009.94,
author = {Zhong, Hao and Zhang, Lu and Xie, Tao and Mei, Hong},
title = {Inferring Resource Specifications from Natural Language API Documentation},
year = {2009},
isbn = {9780769538914},
publisher = {IEEE Computer Society},
address = {USA},
url = {https://doi.org/10.1109/ASE.2009.94},
doi = {10.1109/ASE.2009.94},
booktitle = {Proceedings of the 24th IEEE/ACM International Conference on Automated Software Engineering},
pages = {307–318},
numpages = {12},
series = {ASE '09}
}

@INPROCEEDINGS{7985647,
  author={Zhou, Yu and Gu, Ruihang and Chen, Taolue and Huang, Zhiqiu and Panichella, Sebastiano and Gall, Harald},
  booktitle={2017 IEEE/ACM 39th International Conference on Software Engineering (ICSE)}, 
  title={Analyzing APIs Documentation and Code to Detect Directive Defects}, 
  year={2017},
  volume={},
  number={},
  pages={27-37},
  doi={10.1109/ICSE.2017.11}}

@inproceedings{dinella2022toga,
author = {Dinella, Elizabeth and Ryan, Gabriel and Mytkowicz, Todd and Lahiri, Shuvendu K.},
title = {TOGA: a neural method for test oracle generation},
year = {2022},
isbn = {9781450392211},
publisher = {Association for Computing Machinery},
address = {New York, NY, USA},
url = {https://doi.org/10.1145/3510003.3510141},
doi = {10.1145/3510003.3510141},
booktitle = {Proceedings of the 44th International Conference on Software Engineering},
pages = {2130–2141},
numpages = {12},
location = {Pittsburgh, Pennsylvania},
series = {ICSE '22}
}

@ARTICLE{mastropaolo2023usingtransfer,
  author={Mastropaolo, Antonio and Cooper, Nathan and Palacio, David Nader and Scalabrino, Simone and Poshyvanyk, Denys and Oliveto, Rocco and Bavota, Gabriele},
  journal={IEEE Transactions on Software Engineering}, 
  title={Using Transfer Learning for Code-Related Tasks}, 
  year={2023},
  volume={49},
  number={4},
  pages={1580-1598},
  doi={10.1109/TSE.2022.3183297}}

@inproceedings{10.1145/3524481.3527220,
author = {Tufano, Michele and Drain, Dawn and Svyatkovskiy, Alexey and Sundaresan, Neel},
title = {Generating accurate assert statements for unit test cases using pretrained transformers},
year = {2022},
isbn = {9781450392860},
publisher = {Association for Computing Machinery},
address = {New York, NY, USA},
url = {https://doi.org/10.1145/3524481.3527220},
doi = {10.1145/3524481.3527220},
booktitle = {Proceedings of the 3rd ACM/IEEE International Conference on Automation of Software Test},
pages = {54–64},
numpages = {11},
location = {Pittsburgh, Pennsylvania},
series = {AST '22}
}

@inproceedings{lemieux2023codamosa,
author = {Lemieux, Caroline and Inala, Jeevana Priya and Lahiri, Shuvendu K. and Sen, Siddhartha},
title = {CodaMosa: Escaping Coverage Plateaus in Test Generation with Pre-Trained Large Language Models},
year = {2023},
isbn = {9781665457019},
publisher = {IEEE Press},
url = {https://doi.org/10.1109/ICSE48619.2023.00085},
doi = {10.1109/ICSE48619.2023.00085},
booktitle = {Proceedings of the 45th International Conference on Software Engineering},
pages = {919–931},
numpages = {13},
location = {Melbourne, Victoria, Australia},
series = {ICSE '23}
}

@article{coverup2025,
author = {Altmayer Pizzorno, Juan and Berger, Emery D.},
title = {CoverUp: Effective High Coverage Test Generation for Python},
year = {2025},
issue_date = {July 2025},
publisher = {Association for Computing Machinery},
address = {New York, NY, USA},
volume = {2},
number = {FSE},
url = {https://doi.org/10.1145/3729398},
doi = {10.1145/3729398},
journal = {Proc. ACM Softw. Eng.},
month = jun,
articleno = {FSE128},
numpages = {23}
}

@misc{lahiri2023interactivecodegenerationtestdriven,
      title={Interactive Code Generation via Test-Driven User-Intent Formalization}, 
      author={Shuvendu K. Lahiri and Sarah Fakhoury and Aaditya Naik and Georgios Sakkas and Saikat Chakraborty and Madanlal Musuvathi and Piali Choudhury and Curtis von Veh and Jeevana Priya Inala and Chenglong Wang and Jianfeng Gao},
      year={2023},
      eprint={2208.05950},
      archivePrefix={arXiv},
      primaryClass={cs.SE},
      url={https://arxiv.org/abs/2208.05950}, 
}

@misc{tufano2021unittestcasegeneration,
      title={Unit Test Case Generation with Transformers and Focal Context}, 
      author={Michele Tufano and Dawn Drain and Alexey Svyatkovskiy and Shao Kun Deng and Neel Sundaresan},
      year={2021},
      eprint={2009.05617},
      archivePrefix={arXiv},
      primaryClass={cs.SE},
      url={https://arxiv.org/abs/2009.05617}, 
}

@inproceedings{10.1109/ICSE43902.2021.00112,
author = {Molina, Facundo and Ponzio, Pablo and Aguirre, Nazareno and Frias, Marcelo},
title = {EvoSpex: An Evolutionary Algorithm for Learning Postconditions},
year = {2021},
isbn = {9781450390859},
publisher = {IEEE Press},
url = {https://doi.org/10.1109/ICSE43902.2021.00112},
doi = {10.1109/ICSE43902.2021.00112},
booktitle = {Proceedings of the 43rd International Conference on Software Engineering},
pages = {1223–1235},
numpages = {13},
location = {Madrid, Spain},
series = {ICSE '21}
}

@misc{vikram2024largelanguagemodelswrite,
      title={Can Large Language Models Write Good Property-Based Tests?}, 
      author={Vasudev Vikram and Caroline Lemieux and Joshua Sunshine and Rohan Padhye},
      year={2024},
      eprint={2307.04346},
      archivePrefix={arXiv},
      primaryClass={cs.SE},
      url={https://arxiv.org/abs/2307.04346}, 
}

@inproceedings{
DBLP:journals/corr/abs-2210-00848,
title={Toward Trustworthy Neural Program Synthesis},
author={Wen-Ding Li and Darren Yan Key and Kevin Ellis},
booktitle={ICLR 2025 Third Workshop on Deep Learning for Code},
year={2025},
url={https://openreview.net/forum?id=zC4Wjyu2Wu}
}

@inproceedings{10.1145/2837614.2837664,
author = {Garg, Pranav and Neider, Daniel and Madhusudan, P. and Roth, Dan},
title = {Learning invariants using decision trees and implication counterexamples},
year = {2016},
isbn = {9781450335492},
publisher = {Association for Computing Machinery},
address = {New York, NY, USA},
url = {https://doi.org/10.1145/2837614.2837664},
doi = {10.1145/2837614.2837664},
booktitle = {Proceedings of the 43rd Annual ACM SIGPLAN-SIGACT Symposium on Principles of Programming Languages},
pages = {499–512},
numpages = {14},
location = {St. Petersburg, FL, USA},
series = {POPL '16}
}

@inproceedings{
Laich2020Guiding,
title={Guiding Program Synthesis by Learning to Generate Examples},
author={Larissa Laich and Pavol Bielik and Martin Vechev},
booktitle={International Conference on Learning Representations},
year={2020},
url={https://openreview.net/forum?id=BJl07ySKvS}
}

@inproceedings{10.1145/3385412.3385986,
author = {Yao, Jianan and Ryan, Gabriel and Wong, Justin and Jana, Suman and Gu, Ronghui},
title = {Learning nonlinear loop invariants with gated continuous logic networks},
year = {2020},
isbn = {9781450376136},
publisher = {Association for Computing Machinery},
address = {New York, NY, USA},
url = {https://doi.org/10.1145/3385412.3385986},
doi = {10.1145/3385412.3385986},
booktitle = {Proceedings of the 41st ACM SIGPLAN Conference on Programming Language Design and Implementation},
pages = {106–120},
numpages = {15},
location = {London, UK},
series = {PLDI 2020}
}

@InProceedings{pmlr-v202-pei23a,
  title = 	 {Can Large Language Models Reason about Program Invariants?},
  author =       {Pei, Kexin and Bieber, David and Shi, Kensen and Sutton, Charles and Yin, Pengcheng},
  booktitle = 	 {Proceedings of the 40th International Conference on Machine Learning},
  pages = 	 {27496--27520},
  year = 	 {2023},
  editor = 	 {Krause, Andreas and Brunskill, Emma and Cho, Kyunghyun and Engelhardt, Barbara and Sabato, Sivan and Scarlett, Jonathan},
  volume = 	 {202},
  series = 	 {Proceedings of Machine Learning Research},
  month = 	 {23--29 Jul},
  publisher =    {PMLR},
  url = 	 {https://proceedings.mlr.press/v202/pei23a.html}
}

@misc{hoareprompt,
      title={HoarePrompt: Structural Reasoning About Program Correctness in Natural Language}, 
      author={Dimitrios Stamatios Bouras and Yihan Dai and Tairan Wang and Yingfei Xiong and Sergey Mechtaev},
      year={2025},
      eprint={2503.19599},
      archivePrefix={arXiv},
      primaryClass={cs.SE},
      url={https://arxiv.org/abs/2503.19599}, 
}

@inproceedings{le-etal-2026-specmind,
    title = "{S}pec{M}ind: Cognitively Inspired, Interactive Multi-Turn Framework for Postcondition Inference",
    author = "Le, Cuong Chi  and
      Pham, Minh V.t.  and
      Vu, Tung D.  and
      Cuong, Van Duc  and
      Huy, Phan Nhat  and
      Hoang, Phan Nhat  and
      Nguyen, Tien N.",
    editor = "Liakata, Maria  and
      Moreira, Viviane P.  and
      Zhang, Jiajun  and
      Jurgens, David",
    booktitle = "Proceedings of the 64th Annual Meeting of the {A}ssociation for {C}omputational {L}inguistics (Volume 1: Long Papers)",
    month = jul,
    year = "2026",
    address = "San Diego, California, United States",
    publisher = "Association for Computational Linguistics",
    url = "https://aclanthology.org/2026.acl-long.1687/",
    pages = "36409--36424",
    ISBN = "979-8-89176-390-6"
}

@inproceedings{ruan2025specrover,
author = {Ruan, Haifeng and Zhang, Yuntong and Roychoudhury, Abhik},
title = {SpecRover: Code Intent Extraction via LLMs},
year = {2025},
isbn = {9798331505691},
publisher = {IEEE Press},
url = {https://doi.org/10.1109/ICSE55347.2025.00080},
doi = {10.1109/ICSE55347.2025.00080},
booktitle = {Proceedings of the IEEE/ACM 47th International Conference on Software Engineering},
pages = {963–974},
numpages = {12},
location = {Ottawa, Ontario, Canada},
series = {ICSE '25}
}

@inproceedings{C2,
 author = {Peyman Oreizy and Nenad Medvidovic and Richard N. Taylor},
 title = {Architecture-based runtime software evolution},
 booktitle = {Proceedings of the 20th international conference on Software engineering},
 year = {1998},
 pages = {177--186},
 publisher = {IEEE Computer Society},
}

@inproceedings{zeller07,
 author = {Wasylkowski, Andrzej and Zeller, Andreas and Lindig, Christian},
 title = {Detecting Object Usage Anomalies},
 booktitle = {Proceedings of the Symposium on Foundations of Software Engineering},
 series = {ESEC-FSE '07},
 year = {2007},
 isbn = {978-1-59593-811-4},
 location = {Dubrovnik, Croatia},
 pages = {35--44},
 numpages = {10},
 url = {http://doi.acm.org/10.1145/1287624.1287632},
 doi = {10.1145/1287624.1287632},
 acmid = {1287632},
 publisher = {ACM},
}

@inproceedings{dynamine05,
author = {Livshits, Benjamin and Zimmermann, Thomas},
title = {DynaMine: finding common error patterns by mining software revision histories},
year = {2005},
isbn = {1595930140},
publisher = {Association for Computing Machinery},
address = {New York, NY, USA},
url = {https://doi.org/10.1145/1081706.1081754},
doi = {10.1145/1081706.1081754},
booktitle = {Proceedings of the 10th European Software Engineering Conference Held Jointly with 13th ACM SIGSOFT International Symposium on Foundations of Software Engineering},
pages = {296–305},
numpages = {10},
location = {Lisbon, Portugal},
series = {ESEC/FSE-13}
}

@inproceedings{engler-sosp01,
 author = {Engler, Dawson and Chen, David Yu and Hallem, Seth and Chou, Andy and Chelf, Benjamin},
 title = {Bugs As Deviant Behavior: A General Approach to Inferring Errors in Systems Code},
 booktitle = {Proceedings of the Eighteenth ACM Symposium on Operating Systems Principles},
 series = {SOSP'01},
 year = {2001},
 isbn = {1-58113-389-8},
 location = {Banff, Alberta, Canada},
 pages = {57--72},
 numpages = {16},
 url = {http://doi.acm.org/10.1145/502034.502041},
 doi = {10.1145/502034.502041},
 acmid = {502041},
 publisher = {ACM},
}

@article{williams-tse05,
 author = {Williams,, Chadd C. and Hollingsworth,, Jeffrey K.},
 title = {Automatic Mining of Source Code Repositories to Improve Bug Finding Techniques},
 journal = {IEEE Trans. Softw. Eng.},
 volume = {31},
 number = {6},
 year = {2005},
 pages = {466--480},
 publisher = {IEEE Press},
}

@inproceedings{prminer-fse05,
 author = {Li, Zhenmin and Zhou, Yuanyuan},
 title = {PR-Miner: Automatically Extracting Implicit Programming Rules and Detecting Violations in Large Software Code},
 booktitle = {Proceedings of the 13th Symposium on Foundations of Software Engineering},
 series = {ESEC/FSE-13},
 year = {2005},
 isbn = {1-59593-014-0},
 location = {Lisbon, Portugal},
 pages = {306--315},
 numpages = {10},
 url = {http://doi.acm.org/10.1145/1081706.1081755},
 doi = {10.1145/1081706.1081755},
 acmid = {1081755},
 publisher = {ACM},
}

@inproceedings{ammons-popl02,
 author = {Ammons, Glenn and Bod\'{\i}k, Rastislav and Larus, James R.},
 title = {Mining Specifications},
 booktitle = {Proceedings of the 29th ACM SIGPLAN SIGACT Symposium on Principles of Programming Languages},
 series = {POPL '02},
 year = {2002},
 isbn = {1-58113-450-9},
 location = {Portland, Oregon},
 pages = {4--16},
 numpages = {13},
 url = {http://doi.acm.org/10.1145/503272.503275},
 doi = {10.1145/503272.503275},
 acmid = {503275},
 publisher = {ACM},
}

@INPROCEEDINGS{taoxie-ase09,
 author = {Thummalapenta, Suresh and Xie, Tao},
 title = {Alattin: Mining Alternative Patterns for Detecting Neglected Conditions},
 booktitle = {Proceedings of the 2009 IEEE/ACM International Conference on Automated Software Engineering},
 series = {ASE '09},
 year = {2009},
 isbn = {978-0-7695-3891-4},
 pages = {283--294},
 numpages = {12},
 url = {http://dx.doi.org/10.1109/ASE.2009.72},
 doi = {10.1109/ASE.2009.72},
 acmid = {1747526},
 publisher = {IEEE Computer Society},
}

@INPROCEEDINGS{zeller-ase09,
 author = {Wasylkowski, Andrzej and Zeller, Andreas},
 title = {Mining Temporal Specifications from Object Usage},
 booktitle = {Proceedings of the 2009 IEEE/ACM International Conference on Automated Software Engineering},
 series = {ASE '09},
 year = {2009},
 isbn = {978-0-7695-3891-4},
 pages = {295--306},
 numpages = {12},
 url = {http://dx.doi.org/10.1109/ASE.2009.30},
 doi = {10.1109/ASE.2009.30},
 acmid = {1747527},
 publisher = {IEEE Computer Society},
}

@INPROCEEDINGS{mike-ase09,
 author = {Pradel, Michael and Gross, Thomas R.},
 title = {Automatic Generation of Object Usage Specifications from Large Method Traces},
 booktitle = {Proceedings of the 2009 IEEE/ACM International Conference on Automated Software Engineering},
 series = {ASE '09},
 year = {2009},
 isbn = {978-0-7695-3891-4},
 pages = {371--382},
 numpages = {12},
 url = {http://dx.doi.org/10.1109/ASE.2009.60},
 doi = {10.1109/ASE.2009.60},
 acmid = {1747533},
 publisher = {IEEE Computer Society},
}

@inproceedings{ramanathan-pldi07,
 author = {Ramanathan, Murali Krishna and Grama, Ananth and Jagannathan, Suresh},
 title = {Static Specification Inference Using Predicate Mining},
 booktitle = {Proceedings of the 2007 ACM SIGPLAN Conference on Programming Language Design and Implementation},
 series = {PLDI '07},
 year = {2007},
 isbn = {978-1-59593-633-2},
 location = {San Diego, California, USA},
 pages = {123--134},
 numpages = {12},
 url = {http://doi.acm.org/10.1145/1250734.1250749},
 doi = {10.1145/1250734.1250749},
 acmid = {1250749},
 publisher = {ACM},
}

@inproceedings{fse09,
 author = {Nguyen, Tung Thanh and Nguyen, Hoan Anh and Pham, Nam H. and Al-Kofahi, Jafar M. and Nguyen, Tien N.},
 title = {Graph-based Mining of Multiple Object Usage Patterns},
 booktitle = {Proceedings of the Symposium on Foundations of Software Engineering},
 series = {ESEC/FSE '09},
 year = {2009},
 isbn = {978-1-60558-001-2},
 location = {Amsterdam, The Netherlands},
 pages = {383--392},
 numpages = {10},
 url = {http://doi.acm.org/10.1145/1595696.1595767},
 doi = {10.1145/1595696.1595767},
 acmid = {1595767},
 publisher = {ACM},
}

@inproceedings{mapo09,
 author = {Zhong, Hao and Xie, Tao and Zhang, Lu and Pei, Jian and Mei, Hong},
 title = {MAPO: Mining and Recommending API Usage Patterns},
 booktitle = {Proceedings of the 23rd European Conference on ECOOP 2009 --- Object-Oriented Programming},
 year = {2009},
 pages = {318--343},
 publisher = {Springer-Verlag},
}

@inproceedings{kremenek06,
author = {Kremenek, Ted and Twohey, Paul and Back, Godmar and Ng, Andrew and Engler, Dawson},
title = {From uncertainty to belief: inferring the specification within},
year = {2006},
isbn = {1931971471},
publisher = {USENIX Association},
address = {USA},
booktitle = {Proceedings of the 7th Symposium on Operating Systems Design and Implementation},
pages = {161–176},
numpages = {16},
location = {Seattle, Washington},
series = {OSDI '06}
}

@inproceedings{daikon99,
 author = {Ernst, Michael D. and Cockrell, Jake and Griswold, William G. and Notkin, David},
 title = {Dynamically Discovering Likely Program Invariants to Support Program Evolution},
 booktitle = {Proceedings of the 21st International Conference on Software Engineering},
 series = {ICSE'99},
 year = {1999},
 isbn = {1-58113-074-0},
 location = {Los Angeles, California, USA},
 pages = {213--224},
 numpages = {12},
 url = {http://doi.acm.org/10.1145/302405.302467},
 doi = {10.1145/302405.302467},
 acmid = {302467},
 publisher = {ACM},
}

@inproceedings{yiwei-icse11,
 author = {Wei, Yi and Furia, Carlo A. and Kazmin, Nikolay and Meyer, Bertrand},
 title = {Inferring Better Contracts},
 booktitle = {Proceedings of the 33rd International Conference on Software Engineering},
 series = {ICSE '11},
 year = {2011},
 isbn = {978-1-4503-0445-0},
 location = {Waikiki, Honolulu, HI, USA},
 pages = {191--200},
 numpages = {10},
 url = {http://doi.acm.org/10.1145/1985793.1985820},
 doi = {10.1145/1985793.1985820},
 acmid = {1985820},
 publisher = {ACM},
}

@inproceedings{BeschastnikhBSSE2011,
 author = {Beschastnikh, Ivan and Brun, Yuriy and Schneider, Sigurd and Sloan, Michael and Ernst, Michael D.},
 title = {Leveraging Existing Instrumentation to Automatically Infer Invariant-constrained Models},
 booktitle = {Proceedings of the 19th Symposium on Foundations of Software Engineering},
 series = {ESEC/FSE '11},
 year = {2011},
 isbn = {978-1-4503-0443-6},
 location = {Szeged, Hungary},
 pages = {267--277},
 numpages = {11},
 url = {http://doi.acm.org/10.1145/2025113.2025151},
 doi = {10.1145/2025113.2025151},
 acmid = {2025151},
 publisher = {ACM},
}

@dataset{speccoder-website,
  url          = {https://speccoder.site},
}

\end{document}